\documentclass[fleqn,usenatbib]{mnras}

\usepackage{newtxtext,newtxmath}

\usepackage[T1]{fontenc}

\DeclareRobustCommand{\VAN}[3]{#2}
\let\VANthebibliography\thebibliography
\def\thebibliography{\DeclareRobustCommand{\VAN}[3]{##3}\VANthebibliography}

\usepackage{graphicx}	
\usepackage{amsmath}	

\title[A method to distinguish between micro- and macro-granular surfaces of small Solar System bodies]{A method to distinguish between micro- and macro-granular surfaces of small Solar System bodies}

\author[D. Bischoff et al.]{
D. Bischoff\thanks{E-mail: d.bischoff@tu-bs.de},
B. Gundlach,
J. Blum\\
Institut f\"ur Geophysik und Extraterrestrische Physik, Technische Universit\"at Braunschweig, Mendelssohnstr. 3, 38106 Braunschweig, Germany\\
}

\date{Accepted XXX. Received YYY; in original form ZZZ}

\pubyear{2021}

\begin{document}
\label{firstpage}
\pagerange{\pageref{firstpage}--\pageref{lastpage}}
\maketitle

\begin{abstract}
The surface granularity of small Solar System bodies is diverse through the different types of planetary bodies and even for specific objects it is often not known in detail. One of the physical properties that strongly depends on the surface structure is the surface temperature. In highly porous media with large voids, radiation can efficiently transport heat, whereas more compact, micro-porous structures transport the heat primarily by conduction through the solid material. In this work, we investigate under which conditions a macro-porous surface can be distinguished from a micro-porous one by simply measuring the surface temperature. In our numerical simulations, we included circular and elliptical orbits with and without obliquity and varied the rotation period of the considered objects. We found that daily temperature cycles are rather insensitive to the specific surface granularity. However, the surface temperature at sunrise shows significant dependency on the material structure and this effect becomes even more pronounced when the solar intensity increases. By measuring the sunrise temperature as a function of insolation at noon, a differentiation between micro- and macro-granular surface structures is possible. In this paper, we provide a strategy how remote sensing can be used to derive the surface structure of small Solar System bodies. 
\end{abstract}



\section{Introduction}
\label{sec:introduction}

There are several different types of small bodies in the Solar System, like comets, asteroids, or Kuiper Belt objects, to name just a few. In general, small Solar System bodies are all believed to be structurally granular \citep{Hestroffer.2019}. However, the degree and length scale of granularity can vary. Bodies in the asteroid belt have undergone a long evolutionary history, including compaction \citep{Beitz.2016} and differentiation (large bodies), or catastrophic disruption with possible re-accretion of rubble-pile asteroids (small bodies) \citep{Asphaug.2009, Walsh.2018}. Kuiper Belt objects likely formed by the collapse of dust clouds consisting of mm- to dm sized pebbles \citep{Robinson.2020,McKinnon.2020}, which produced objects with an initial granularity. Comets either originate from the low-mass tail of the initial planetesimal mass function or are remnants of the collisional evolution in the Kuiper Belt \citep[][]{Weissman.2020}.
\par 
The thermal observation of small Solar System bodies started from ground-based facilities, but their limited spatial resolution resulted in global mean values with high deviations. In the past years, several space missions to small bodies were equipped with instruments able to measure surface temperatures with high spatial resolution, partly also in the sub-surface regions. This data allows a detailed look on the thermal properties and on the applicability of theoretical thermophysical models. For comets 1P/Halley \citep{Emerich.1988}, 9P/Tempel 1 \citep{Groussin.2007}, 19P/Borrelly \citep{Soderblom.2004} and 103P/Hartley 2 \citep{Groussin.2013}, temperature data were measured, but only for limited periods in time, due to the fly-by nature of the respective space missions. This changed with the Rosetta mission, which had the instruments VIRTIS \citep[Visible InfraRed and Thermal Imaging Spectrometer,][]{Coradini.2007}, and MIRO \citep[Microwave Instrument for the Rosetta Orbiter,][]{Gulkis.2007} on board the orbiter. These instruments were capable of measuring temperatures of comet 67P/Churyumov-Gerasimenko (hereafter comet 67P) for a long time period during which the comet passed perihelion. Due to the different wavelengths used, VIRTIS and MIRO probed different depths. Additionally, the lander Philae was equipped with the MUPUS instrument \citep[Multipurpose Sensors for Surface and Sub-Surface Science,][]{Spohn.2015}, which included a thermal probe, MUPUS PEN, with 16 resistance temperature detectors and an infrared radiometer, MUPUS TM. Due to the complexity of the landing site, the interpretation of their data can be challenging. However, \citet{Spohn.2015} found a best fitting thermal inertia of $85 \, \mathrm{J/(K \, m^2 \, s^{0.5})}$.
\par
The VIRTIS instrument was capable to measure near-surface temperatures at depths in the range of tens of micrometres, but with a lower temperature detection limit of roughly 156~K, due to instrument noise. Recent results are presented in \citet[][]{Tosi.2019}, showing the effects of self-heating and shadowing. \citet[][]{Tosi.2019} also compared a thermophysical model to the VIRTIS and MIRO data and showed that an ice-free model fits the daytime temperature further away from the Sun, whereas nearer to the Sun, ice needs to be included to match the observations. 
\par The MIRO instrument measured at much longer wavelengths of $1.6\, \mathrm{mm}$ and $0.5\,\mathrm{mm}$, respectively, for which the penetration depth is on the order of a few centimetres \citep{Blum.2017}. As the diurnal skin depth is comparable to this depth \citep{Gulkis.2015,Schloerb.2015}, the radiation MIRO receives has contributions from very different temperatures and is thus difficult to interpret. \citet{Marshall.2018} used MIRO and VIRTIS data to investigate physical properties of the surface and sub-surface of 67P, namely the thermal inertia and roughness of specific locations. For the thermal inertia, they find a best-fitting value of $80 \, \mathrm{J/(K \, m^2 \, s^{0.5})}$ and a complex roughness distribution. It should be noted here that in those cases, in which radiative heat transfer is of importance, the thermal inertia should be temperature dependent so that a single value has a limited meaning.
\par 
On the Dawn spacecraft, which visited the asteroids Vesta and Ceres, the VIRTIS instrument was also installed. A map of the surface thermal properties of the asteroid Vesta was generated by \citet{Capria.2014}. They argue that the surface of Vesta is covered by a fine regolith, due to the low thermal inertia of roughly $30 \, \mathrm{J/(K \, m^2 \, s^{0.5})}$, but they also find varying surface properties. Due to the usage of the same instrument, the same restriction for lower temperatures also apply here.
\par 
Two near-Earth asteroids were recently visited by the space missions OSIRIS-REx and Hayabusa 2. OSIRIS-REx observed and took a sample from (101955) Bennu \citep{Lauretta.2019}, whereas the Hayabusa 2 was sent to (162173) Ryugu, also taking a sample, which finally arrived on Earth in December 2020 \citep{Watanabe.2019b}. Both bodies are believed to be rubble-pile asteroids and show a top-shape, typical for fast rotating asteroids, which accumulated material at the equator \citep{Hirabayashi.2020}, but which can also be explained by direct formation due to re-accumulation after disruption \citep{Michel.2020}. Due to the relatively low thermal inertia, deduced from telescopic and spatially unresolved infrared observations, a mainly regolith-covered surface was expected for Ryugu \citep{Wada.2018}. Thus, the finding of large boulders all over the surface was surprising. Our new model described in this paper potentially offers a method to distinguish between a regolith and a boulder covered surface in future thermal modelling. Observations of the boulder size-frequency distribution on Bennu indicate, together with the moderate thermal inertia, that the dependency of the thermal inertia on particle size is more complex than previously assumed \citep[][]{DellaGiustina.2019}. They found a global thermal inertia of $(350\pm 20)  \, \mathrm{J/(K \, m^2  \, s^{-0.5})}$. \citet{Rozitis.2020} modelled the surface and sub-surface temperatures of Bennu to address the sublimation of water ice and thermal fracturing. They found sufficiently cold regions for near-surface water ice and high daily temperature variations in warmer regions, which can lead to rock fractures and, in turn, to particle ejection. As part of the Hayabusa 2 mission, MASCOT landed on Ryugu and the instrument MARA directly measured the temperature of a nearby boulder. From this measurement, a diurnal cycle of surface temperatures of this boulder was followed and analysed by \citet{Grott.2019} and \citet{Hamm.2020}. The model fit of \citet{Grott.2019} to the MARA data indicates a thermal inertia of $282^{+93}_{-35} \, \mathrm{J/(K \, m^2  \, s^{-0.5})}$, equivalent to a heat conductivity of $\sim 0.1 \, \mathrm{W/(K \, m)}$ and a high porosity of $28\%$ to $55 \%$. They also describe the highly variable daytime temperatures, which could not be fitted by a simple model. However, it was shown that a model including surface roughness matches better. For explaining the nighttime temperatures, the implementation of roughness was not needed, indicating that roughness and other influential effects, like shadowing or self-heating seem to be negligible at night. \citet{Hamm.2020} used the data assimilation method to retrieve thermophysical properties of the observed boulder and estimated a thermal inertia of $(295\pm 18)  \, \mathrm{J/(K \, m^2  \, s^{-0.5})}$, equivalent to a heat conductivity in the range of $\sim 0.07 - 0.12 \, \mathrm{W/(K \, m)}$. The porosity is expected to be between $30\%$ to $52 \%$, matching the findings of \citet{Grott.2019}. Fitting a constant thermal inertia to the global temperature data of Ryugu results in comparable ranges as the analysis of the boulder \citep[][]{Sugita.2019, Okada.2020,Shimaki.2020}, hinting to a generally higher porosity and less consolidation than expected. Additionally, \citet{Sakatani.2021} found boulders with a porosity of $\geq 70 \%$, which are expected to be the most primordial material on Ryugu. Micro- and macro-porosity of Ryugu was addressed by \citet{Grott.2020}. They found a smaller than previously assumed macro-porosity of $16\%\pm 3\%$ influenced by the polydisperse particles, leading, together with the high micro-porosity of $\sim 50\%$, to an average grain density of $\sim 2850 \, \mathrm{kg/m^3}$.

\par 
In this work, we concentrate on the influence of the granularity of the surfaces of small Solar System bodies on their thermal properties and therewith on their surface temperatures. As shown in previous works \citep[][]{Gundlach.2012,Gundlach.2013,Gundlach.2020}, the length-scale of granularity and the porosity determines which heat-transport mechanism, namely conduction, radiation, or gas diffusion dominates. Radiation is only effective in void spaces so that a larger radiative mean free path results in a higher radiative heat conductivity. However, a larger length-scale of the granularity also often results in smaller contact areas between the solid particles per unit area, which hinders the heat flow by conduction. In compact materials, radiation is completely suppressed and can be neglected, but heat conduction through the material is more efficient. In this work, we neglect the effect of gas conductivity in pore spaces, due to the typically very small gas densities in airless bodies. The key difference between the radiative and conductive heat-transport mechanisms is their inherent temperature dependency. While conduction typically is only slightly temperature-dependent, the radiative heat conductivity $\lambda_{\mathrm{rad}}\propto T^3$ depends strongly on temperature $T$. The aim of this work is to investigate whether these different temperature dependencies of the above mentioned heat transport processes can lead to a measurable change of the expected surface temperatures and whether this effect can be used to draw conclusions with respect to the surface structure of the observed objects.
\par
Previous studies investigated the radiative heat transport in detail. \citet{Ryan.2020} focused on the influence of polydisperse particles, including non-isothermality, and found that the temperature dependency of the radiative thermal conductivity can be less than cubic for small material conductivities. In addition, they showed that a power-law particle size-frequency distribution can be described by a monodisperse material with the Sauter mean diameter, calculated as the average of the volume-to-surface ratio of the grains. Hence, it is sufficient to study the influence of radiation and conduction on the heat transport efficiency for monodisperse media.
\par
Our own previous work on heat transport through granular materials showed via the calibration to lunar regolith that irregular polydisperse grains can be approximated by monodisperse spherical grains when introducing a factor $\xi$, which reduces the heat conductivity \citep[][see their Eq. 4 and 5]{Gundlach.2013}. Furthermore, we demonstrated that radiative heat transport plays an important role in the surface layers of comet 67P \citep{Blum.2017,Gundlach.2020} and that, considering the presence of water and carbon-dioxide ice, the emission of water-ice containing dust chunks at the observed size and rate was explicable \citep[][]{Gundlach.2020}.
\par
This paper is structured in the following way: in Section \ref{sec:Formation_scenarios_surface_structures}, the possible processes that lead to either granular or compact surfaces on small Solar System bodies are described, which would result in different thermal properties. In addition, their connections to the different types of small bodies are discussed, according to proposed formation and evolution scenarios. Section \ref{sec:strategy} outlines our strategy to investigate the thermal variation on the surface of small bodies and the influence of thermal radiation on heat transport. The thermophysical model is described in Section \ref{sec:thermo_modell} and the simulation results are presented in Section \ref{sec:results}. We conclude and summarise our findings in Section \ref{sec:conclusions}. Our outlook in Section \ref{sec:outlook} addresses open aspects regarding our model assumptions, on which further work is needed.

\section{Formation Of Surface Structures}
\label{sec:Formation_scenarios_surface_structures}

Several processes lead to the formation, or alteration, of specific morphological structures, especially at the surface of a small body in the Solar System. In the following, the mechanisms that lead to granular structures on different length scales are described. 
\par
A widely discussed formation scenario of planetesimals is the gravoturbulent collapse of a pebble cloud, which was concentrated by, e.g., the streaming instability in the protoplanetary disc \citep[][]{Johansen.2006}. The mm- to cm-sized pebbles themselves were formed by hit-and-stick collisions and compacted in the bouncing regime to porosities of typically 60\% \citep[][]{Guttler.2010,Zsom.2010,Lorek.2018}. For small bodies, the gravoturbulent collapse was so gentle that the pebbles remained intact \citep[][]{Skorov.2012,WahlbergJansson.2017} so that the porosity of the pebble assemblage was about 40\% \citep[][]{ORourke.2020}. For large bodies, the mm- to cm-sized granularity was already lost during formation, due to either too high collapse velocity or internal gravitational stress. Bodies formed this way should only exhibit granularity in the micrometre scale. 
\par 
During the subsequent evolution of the planetesimals, impacts may have compacted or destroyed these bodies \citep[][]{Beitz.2016}. Collisions among equal-planetesimals predominantly resulted in fragmentation. The re-accretion of the fragments into new objects is a common explanation of the formation of top-shape rubble-pile asteroids. If the fragments themselves were consolidated enough, a macro-porosity remains, whose length scale is determined by the fragment sizes \citep[][]{Walsh.2018}. Impacts by small projectiles may locally compact the underlying material, thus destroying any macro- or even micro-porosity, and spread impact fragments over the whole surface, which again leads to the formation of macro-porosity \citep[][]{Beitz.2016}
\par 
Another process forming granular structures is the sublimation of volatiles from the surface of an ice-bearing body, in particular cometary nuclei, leaving a refractory and porous matrix behind. The stability of such a structure certainly depends on the material properties and sublimation circumstances. For sublimation pressures higher than the tensile strength of the refractory material, the surface material will most likely be expelled rather than remaining attached to the surface \citep[][]{Kuhrt.1994,Skorov.2012,Blum.2014,Blum.2017,Gundlach.2020}. However, even in this case, some of the emitted dusty material may fall back to the surface and will cover it so that a macro-porous granular surface forms, as witnessed at comet 67P \citep[][]{Hasselmann.2019,Marschall.2020c}.
\par
Finally, most of the small bodies in the inner Solar System are covered by regolith. Even in the case of an originally consolidated surface, the continuous bombardment of the surface by meteoroids causes cratering and leads to the ejection of material. Depending on the size of the target body, its gravity might re-capture the slowest ejecta, which are typically the larger ones of the size distribution, because the heavier ejecta typically obtain lower ejection velocities after impact \citep[][]{Fujiwara.1980,Nakamura.1991,Nakamura.1993,Nakamura.1994,Vickery.1986,Vickery.1987}. Thus, one might expect small bodies with lower gravitational potential to recollect only the largest fragments, whereas large bodies may be covered in finer grains. \citet[][]{Gundlach.2013} have shown that the dominant regolith size can span a range between $\sim 10$ $\mathrm{\mu m}$ and $\sim 1$ cm.
\par
In conclusion, the granularity of the surface material of a small Solar System bodies can appear on different length scales. This will necessarily lead to different heat conductivities and, thus, to different surface temperatures. In this paper, we will derive a method to distinguish between micro- and macro-granular surface materials. In the former case, the heat transfer is dominated by thermal conductivity through the solid network of grains, whereas for macro-granular surfaces, radiative heat transfer dominates, at least during daytime. Applications of our method can be the derivation of surface-grain sizes and setting constraints on the formation processes of the bodies or their surfaces.

\section{Strategy}
\label{sec:strategy}

This work focuses on the surface temperature as an observable of space missions to small bodies in the Solar System. The surface temperature mainly underlies two cycles, the diurnal variation, due to the rotation of the body around its spin axis, and the orbital variation, due to the motion of the body around the Sun. The latter includes seasonal effects, which occur when the spin axis possesses a finite obliquity, i.e. is tilted against the axis of angular momentum of the orbital motion. In principle, for a body with fixed physical properties, the solar illumination determines the surface temperature. 
\par 
To distinguish between micro- and macro-granular surfaces, we assume two main processes of heat transport into the sub-surface layers of a small body. For large grains, thermal radiation can be a very effective means of energy transport in the voids between the grains. In this case, the resulting heat transport is highly temperature dependent. In the other extreme, i.e. for very small grains, the radiative heat flux can be neglected against the solid-state heat conductivity resulting in no or only weak temperature dependency of the thermal conductivity. The implementation of these scenarios into the thermophysical model is described in Section \ref{sec:thermo_modell}. It should be noted here that the building blocks in the macro-porous case possess an internal porosity. If the size of these building blocks is comparable to the thermal skin depth, e.g. for fast rotators, the internal porosity  dominates the heat transport. In this case, the radiative heat transport inside the building blocks can be neglected in comparison to the network conductivity, which implies that a distinction between regolith and consolidated building blocks is impossible.
\par 
Regarding the orbital motion of the body, we will start with the idealised case of a circular orbit to investigate the surface temperature cycles of both model cases, i.e. macro- versus micro-porosity of the surface material. This will include the influence of the solar intensity on heliocentric distance and latitude variations. To address more realistic scenarios, we also considered elliptical orbits, with and without obliquity. The influence of the rotation period is also being investigated.

\section{Thermophysical Model}
\label{sec:thermo_modell}

The thermophysical model used in this work is based on the model developed by \citet{Gundlach.2012} and \citet{Gundlach.2020}. As mentioned in the previous Sections, we assume that the total heat conductivity consists of two parts, namely phononic heat transport through the network of solid grains, for simplicity assumed to be temperature-independent, and photonic heat transport by radiation through the void spaces. Thus, we describe the total heat conductivity $\lambda(T)$ as the sum of the phononic heat conductivity, $\lambda_{\mathrm{net}}$, and the photonic heat conductivity, $\lambda_{\mathrm{rad}}$, 
\begin{equation}
    \lambda(T) = \lambda_{\mathrm{net}} + \lambda_{\mathrm{rad}}(T) .
\end{equation}
In general, the network conduction depends also on temperature, due to the temperature dependency of the material parameters. In our scenario, however, we will neglect this, because this temperature dependency in general is not as strong as that of the radiative part. For example, \citet[][]{Opeil.2020} measured thermal conductivities of five CM chondrites and found an increase of conductivity by not more than a factor of two for a temperature increase from $100 \, \mathrm{K}$ to $300 \, \mathrm{K}$ (see their Fig. 3). Additionally, for some rocky materials even a decreasing heat conductivity is observed for increasing temperatures \citep[][]{Miao.2014,Horai.1971}. 
\par
For model I, which describes the micro-granular surface, the heat conductivity through the network of minuscule grains with radii $r$ is modelled to be
\begin{equation}
    \lambda_{\mathrm{I}} = \lambda_{\mathrm{net}}(r) = \mathrm{constant},
\end{equation}
which we also assume not to depend on depth under the surface. This assumption could be too restrictive for some surfaces. However, to address the depth-dependency of density and volume filling factor, we chose to run our model for the two porosity extremes. The default case is defined by a volume filling factor of the packing of $\Phi_\mathrm{pack} = 0.55$, but we also used $\Phi_\mathrm{pack} = 0.2$ for comparison (see. Fig. \ref{fig:comparison_porosity}). We treat the constant heat conductivity as a free parameter in the simulation runs and vary it over a wide range of values, $\lambda_{\mathrm{I}} = 0.001-0.1 \, \mathrm{W~m^{-1}~K^{-1}}$, to account for a wide range of grain sizes and grain materials.
\par
\begin{figure}
    \centering
    \includegraphics[width=\columnwidth]{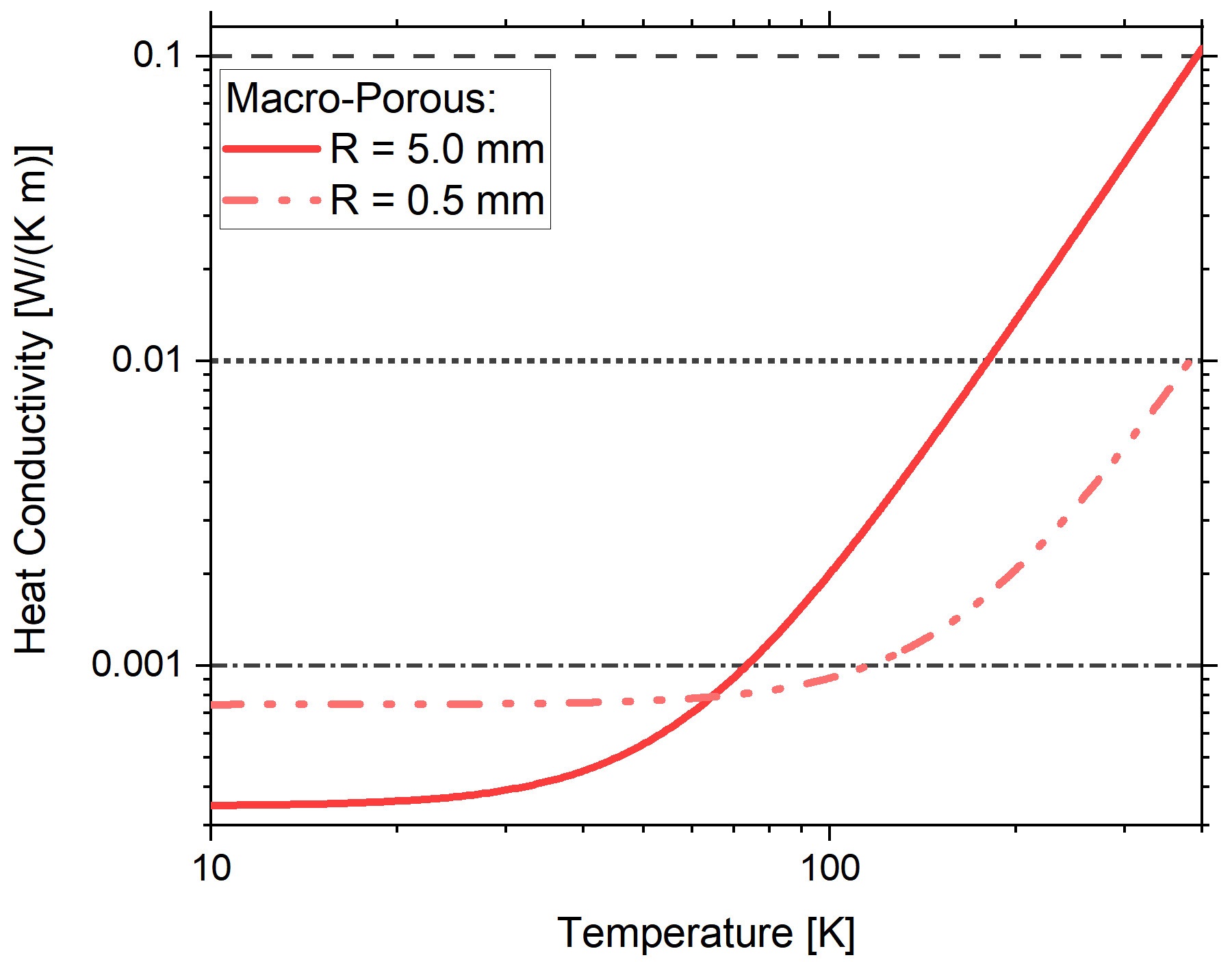}
    \caption{Heat conductivity as a function of temperature for model II. The dashed-dotted line shows the case for pebble radii of $R = 0.5\, \mathrm{mm}$ (light red), whereas the solid curve shows results for pebble radii of $R = 5\, \mathrm{mm}$ (red). For comparison, three constant heat conductivities of model I are highlighted by the three horizontal lines.}
    \label{fig:Lambda_Temperature_Pebble}
\end{figure}

In model II, which aims at describing the macro-granular surface, we assume that the material consists of macroscopic pebbles, for which both, a radiative and a conductive heat transport occurs. Thus, the heat conductivity in this case depends on the assumed pebble radius $R$ and on temperature $T$ and can be written as
\begin{equation}
    \lambda_{\mathrm{II}} = \lambda_{\mathrm{net}}(R) + \lambda_{\mathrm{rad}}(R,T) = \lambda_{\mathrm{net}}(R) + p_{\mathrm{rad}}(R)\, \left( \frac{T}{\mathrm{1~K}} \right)^3. 
    \label{eq:pebble_conductivity}
\end{equation}
The derivation of the parameters
\begin{equation}
    \lambda_{\mathrm{net}}(R) = 5.9 \times 10^{-5} \, \mathrm{\frac{W}{K~m}} \times \left( \frac{R}{\mathrm{1~m}} \right) ^{-1/3}
    \label{eq:pebble:net}
\end{equation}
and
\begin{equation}
    p_{\mathrm{rad}}(R) = 3.3 \times 10^{-7} \, \mathrm{\frac{W}{K~m}} \times \left( \frac{R}{\mathrm{1~m}} \right)
    \label{eq:pebble:rad}
\end{equation}
can be found in Appendix \ref{App:thermophysical_model}, which is based on parameters valid for comet 67P.
Comparing this description with Eqs. \ref{eq:pebble:net} and \ref{eq:pebble:rad} to the measurement by \citet{Sakatani.2018} of the thermal conductivity of the lunar regolith simulant JSC-1A under vacuum conditions shows that in their temperature range of $\sim 250 - 340 \, \mathrm{K}$ our model results in similar values (assuming $R = 0.1 \, \mathrm{mm}$ and without adapting other parameters) with a deviation of $\lesssim 5\%$. For smaller temperatures, our model results in up to $20\%$ lower thermal conductivities than their fitted function.
Due to its strong temperature dependency, the radiative heat conductivity varies over time and depth, according to the temporal and spatial temperature changes. The temperature dependency is illustrated in Fig. \ref{fig:Lambda_Temperature_Pebble} where the thermal conductivity is plotted as a function of temperature for two pebble radii, $R = 5 \, \mathrm{mm}$ (solid red line) and $R = 0.5 \, \mathrm{mm}$ (dashed-dotted light-red line), respectively. For small pebble radii and for microscopic grains, the temperature dependency becomes less relevant and approaches a constant value. For very low temperatures, the temperature dependency can be neglected. The three horizontal lines indicate three constant heat conductivities of model I, which we will use in most of the following simulation runs. From Fig. \ref{fig:Lambda_Temperature_Pebble}, one can see that the variation in heat conductivity in the macro-prosity case can be more than two orders of magnitude over a temperature range of $T\approx 40-400 \, \mathrm{K}$.
In model II, we also do not include depth-dependent material properties, which could be too restrictive. However, for small bodies like asteroids and comets, no stratification of the material was directly measured. The constraints by \citet[][]{Schrapler.2015} show a variation of the volume filling factor from $\sim 0.2$ at the surface to $\sim 0.6$ at depth of tens of centimeters for asteroids (their Fig. 8), and for comets the stratification is unknown. However, we chose, as described for model I, a default volume filling factor of $\Phi_\mathrm{pack} = 0.55$, which corresponds to the expected porosity in the interior. For comparison, we also ran models I and II for $\Phi_\mathrm{pack} = 0.2$, corresponding to the surface expectation.

We describe the temperature stratification and temporal temperature evolution of each unit-area surface element of the small Solar System body by the one-dimensional heat-transfer equation
\begin{equation}
    \rho c \frac{dT(x,t)}{dt} = \frac{d}{dx} \left[ \lambda(T) \frac{dT(x,t)}{dx} \right],
\end{equation}
with the (constant) mass density $\rho$, the temperature $T(x,t)$, depth under the surface $x$, time $t$, heat capacity $c = c_0 \times \left( \frac{T}{\mathrm{200K}} \right)$, and heat conductivity $\lambda(T)$, respectively. A compilation of all parameter values can be found in Table \ref{Tab:1_Parameters}. In this paper, we neglect internal heat sources or losses by, e.g., latent heat of condensation, evaporation or other phase change. Thus, our model is only applicable to ice-free conditions as discussed in Section \ref{sec:outlook}. We assume that the inner temperature of the body equals that of a fast rotator on the same orbit via the following relation
\begin{equation}
    T= \left( \frac{ I_\mathrm{E} (1-A) }{r_\mathrm{h}^2 \, 4 \, \sigma \epsilon} \right)  ^{1/4} .
    \label{eq:eqtemp}
\end{equation}
Here, $I_\mathrm{E} = 1,367 \, \mathrm{W/m^2}$ denotes the solar constant, $A$ the albedo, $r_\mathrm{h}$ the heliocentric distance, $\sigma$ the Stefan-Boltzmann constant and $\epsilon$ the emissivity. The used values can be found in Table \ref{Tab:1_Parameters}. However, the choice of the inner temperature is of minor importance for the rapid diurnal or orbital changes of the surface temperature. To confirm this, we also ran simulations with an inner temperature of $50 \, \mathrm{K}$ and were able to proof this expectation.
\par
In principle, the maximum depth for which a temperature change is calculated, is not fixed and adapts automatically to the energy flow inside. However, due to limitations in computing time, this feature was turned off for elliptical orbits, for which we set the maximum depth to roughly $5 \, \mathrm{m}$, with the boundary condition that the temperature equals that of a fast rotator in the interior.
\par
The heat-transfer equation is solved numerically for discrete depth and time steps, $dx$ and $dt$, respectively. The numerical accuracy of the solution can be described by the Fourier number $F$ \citep{Hensen.1994}, defined as 
\begin{equation}
    F = \frac{\lambda}{\rho c} \frac{dt}{dx^2} .
\end{equation}
For stable numerical conditions, the Fourier number should be $F \leq 0.5$. In case of higher values, the calculations become unstable. For values $F \ll 0.5$, the accuracy of the simulation results decreases. Due to its dependency on the heat conductivity, which may vary over time and depth, the Fourier number is not a constant value in one simulation run. To address this variation, we allowed a range from $F=0.01$ to $F=0.5$ and adapted the depth step $dx$ for each simulation run such that the maximal Fourier number stays in this range. For easier comparison, the time step $dt$ is fixed for a specific set of simulations (e.g., for all simulations on circular orbits). 
\par
With this thermophysical model, the spatial and temporal temperature variations were calculated for several scenarios, taking into account elliptical orbits and obliquity. This enabled us to look at the orbital temperature evolution of an arbitrary surface element. To be independent of the initial condition of depth-independent temperature according to Eq. \ref{eq:eqtemp}, a simulation run covers $1.5$ orbits when considering elliptical orbits and 1,000 rotation periods when considering circular orbits. This is sufficiently long, as the evolution of the surface temperature mostly relies on the instantaneous insolation, as mentioned before, rather than on the internal temperatures, which evolve on long time scales. For simplicity, we assume a spherical body for which only the local longitude and latitude determine the amount of received energy.

\section{Simulation Results}
\label{sec:results}

In this Section, we present the findings from our simulations. To validate our numerical model, we start with a comparison of our model results to measurements and published model data of the asteroid Ryugu, as provided by \citet{Grott.2019}. Thereafter, several simulation scenarios for macro- and micro-porosity cases are presented for an idealised and a more realistic scenario. For the more realistic case, we chose the orbital parameters of comet 67P.

\subsection{Model Validation}
\label{sec:model_validation}
\begin{figure}
    \centering
    \includegraphics[width=\columnwidth]{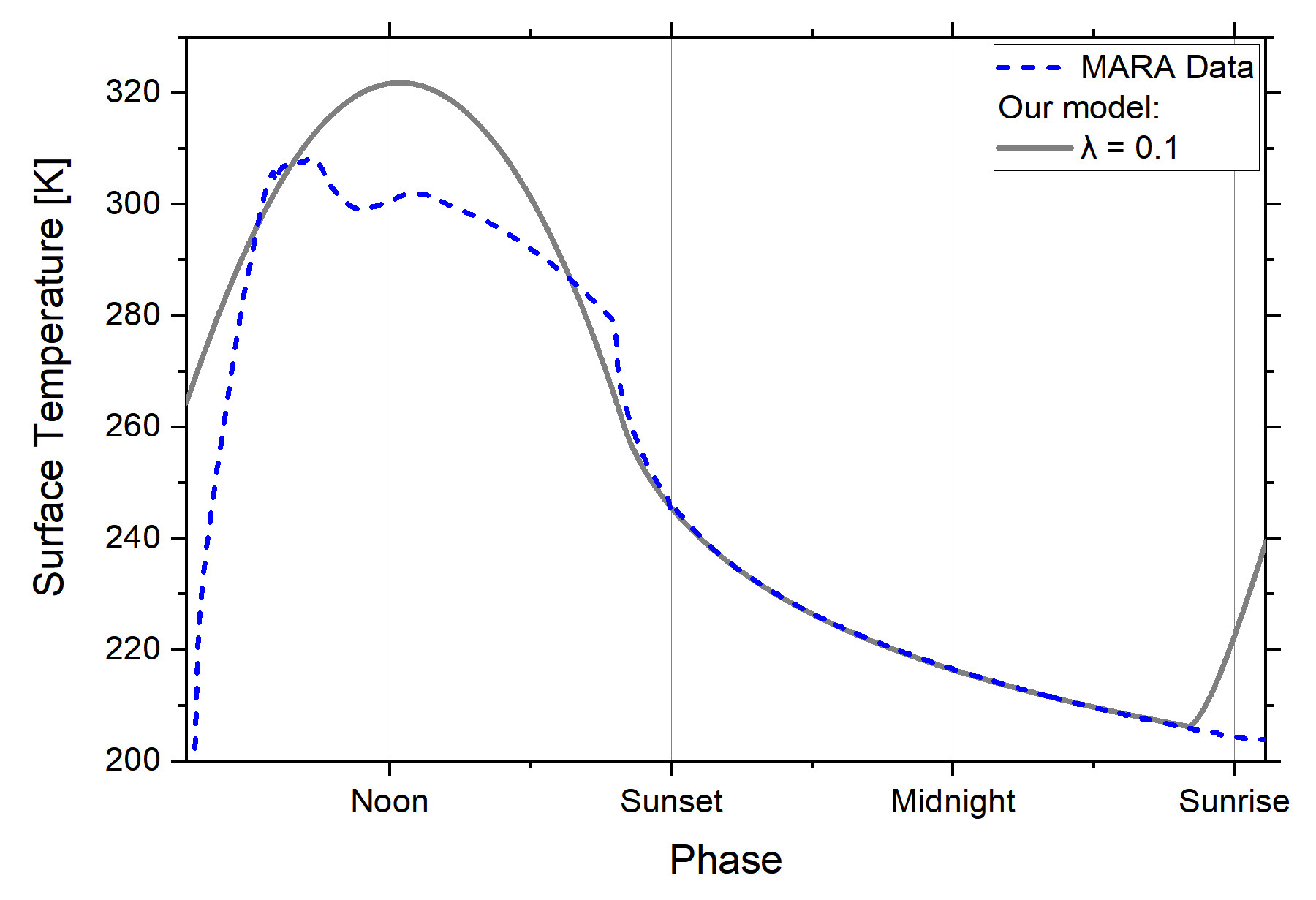}
    \caption{Surface temperature of a boulder on asteroid Ryugu measured by the MARA instrument on board the MASCOT lander \citep[blue dashed curve;][]{Grott.2019} compared to the result of our thermophysical micro-granular model with $\lambda = 0.1\, \mathrm{W/(K \, m)}$ (grey solid curve).}
    \label{fig:Ryugu_comparison}
\end{figure}

To validate our numerical approach and assumptions, we compare our model results to the measurement data published by \citet{Grott.2019}. As described in Section \ref{sec:introduction}, \citet{Grott.2019} and \citet{Hamm.2020} compared their thermophysical model to the surface temperatures of a boulder on asteroid Ryugu, measured by the MARA instrument on board the MASCOT lander. We adjusted our orbital and material parameters to the Ryugu case and calculated the surface temperature evolution. Fig. \ref{fig:Ryugu_comparison} shows the original measured data (blue dashed curve) and our model results (grey solid curve) for the micro-granular case, i.e. for a temperature-independent heat conductivity. For $\lambda = 0.1\, \mathrm{W/(K \, m)}$, our model matches the measured nighttime data of asteroid Ryugu very well. At daytime, the agreement between our model and the Mara measurements is less perfect, but as described by \citet{Grott.2019}, there are subtle illumination details during the daytime hours that a simple model cannot reproduce. Since our one-dimensional model is not capable to simulate complex surface structures, shadowing and other surface effects cannot be taken into account. \citet{Grott.2019} and \citet{Hamm.2020} used the surface orientation of the spot as a free parameter and were able to constrain this parameter even through the nighttime temperatures. We assume a spherical body so that sunrise and sunset are half a rotation period apart. Thus, our simple model cannot explain object-specific surface-orientation effects, as, e.g. responsible for the obvious earlier sunset at Ryugu shown in Fig \ref{fig:Ryugu_comparison}. However, \citet{Grott.2019} and \citet{Hamm.2020} showed that also a non-inclined (i.e., normal) surface orientation can reproduce the nighttime temperatures so that a comparison with our model is feasible, which is also demonstrated by the fact that our model result fall into their thermal-conductivity range. This fact and the agreement between our model and the MARA measurements for the nighttime temperatures validates the general applicability of our numerical implementation of this thermophysical model.

\subsection{The Idealised Case}
\label{sec:results_sub:ideal}

\begin{figure*}
    \begin{minipage}[b]{.47\linewidth}
    \centering
    \includegraphics[width=\columnwidth]{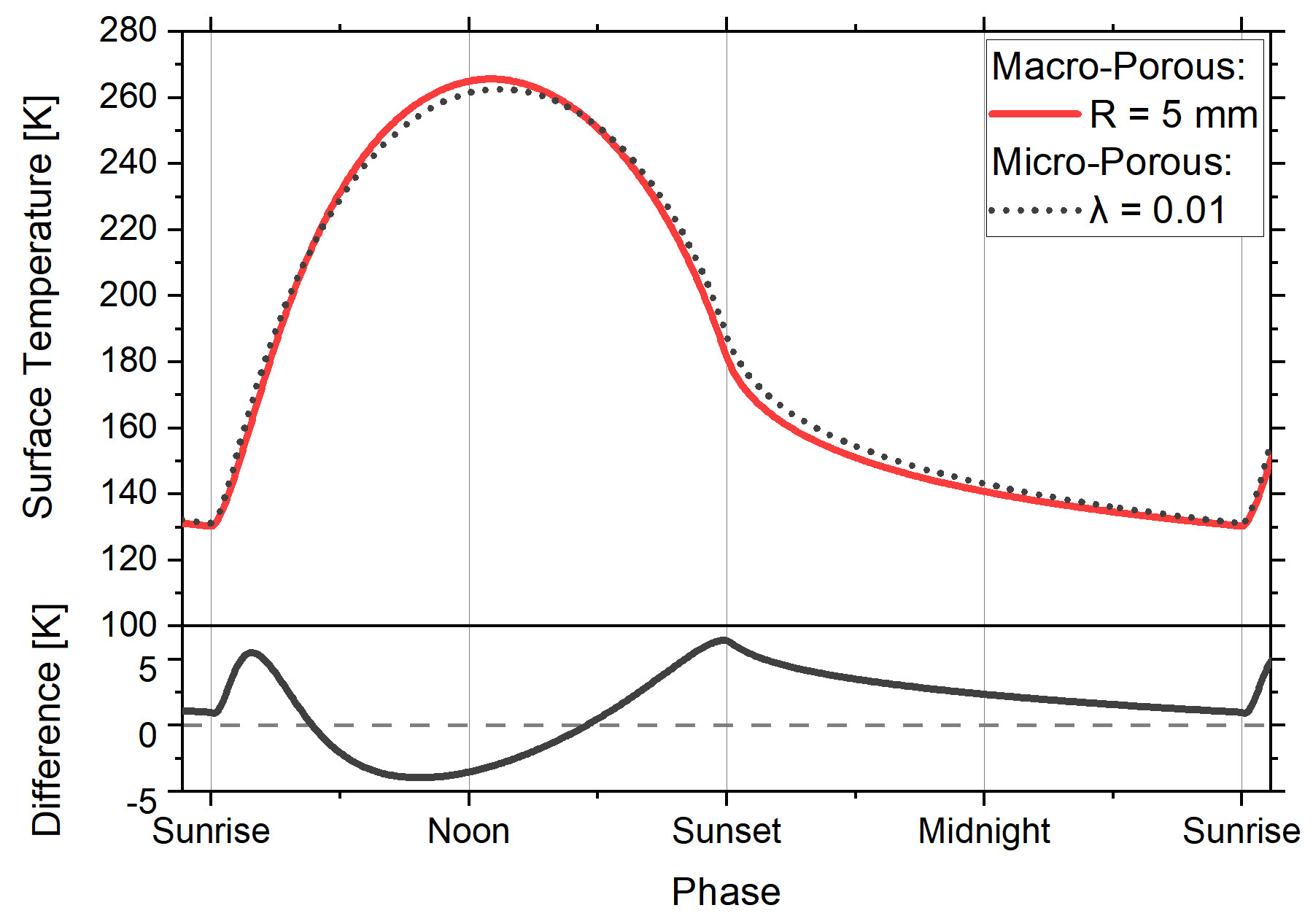}
    \end{minipage}
    \hspace{.05\linewidth}
    \begin{minipage}[b]{.47\linewidth}
    \includegraphics[width=\columnwidth]{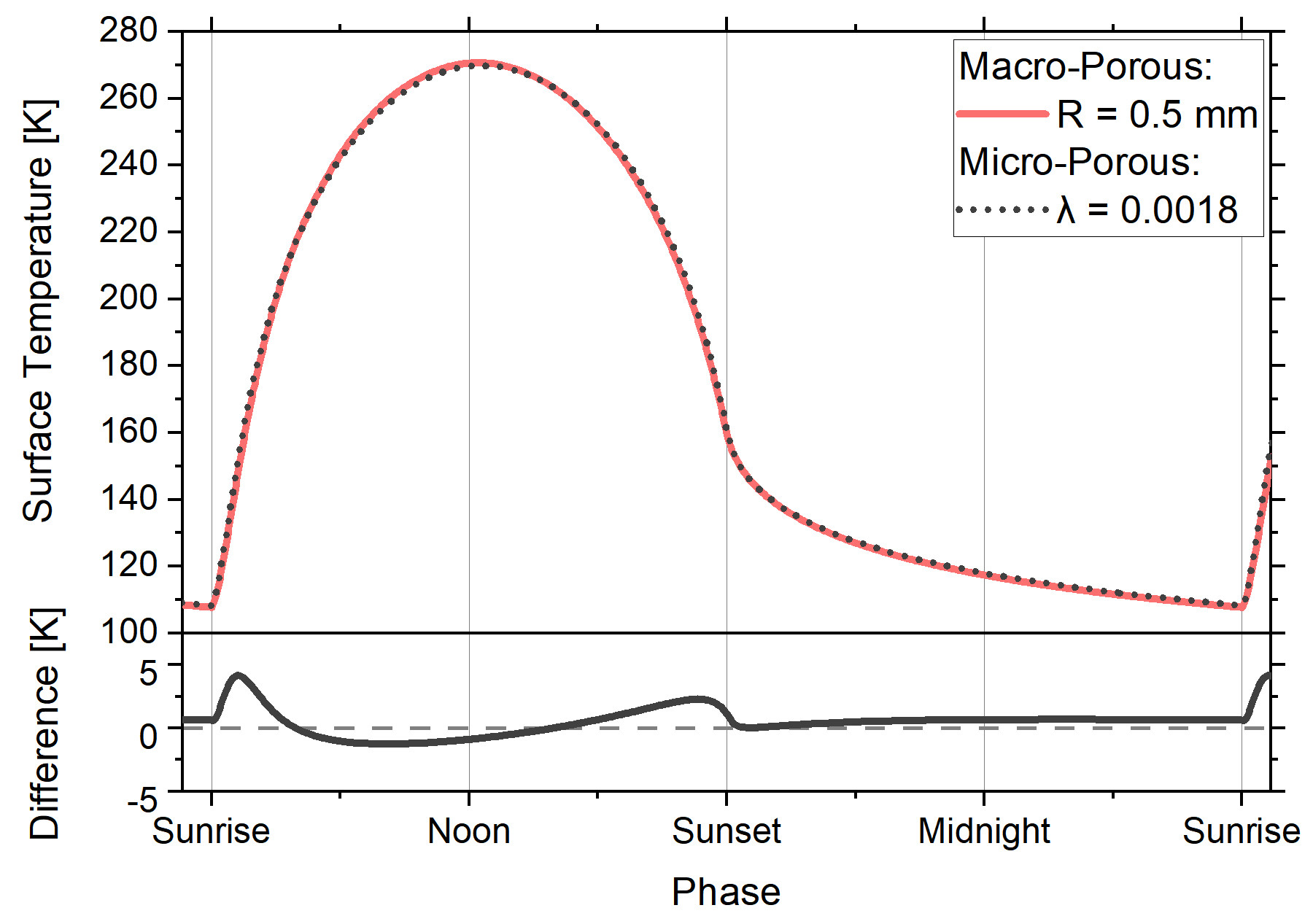}
    \end{minipage}
    \caption{Diurnal evolution of the equatorial surface temperature for the macro-porosity case (red solid curves), after 998 days of simulation. The simulation was run for a circular orbit without obliquity and at a heliocentric distance of $r_H = 2  \, \mathrm{AU}$. For comparison, the best-fitting micro-porosity case is shown by the respective grey dotted curves. Below the diurnal temperature profiles, the temperature difference between the macro- and the micro-porosity case is shown. Left: simulation result for a pebble radius of $R = 5 \, \mathrm{mm}$ and a micro-porosity case with a temperature-independent heat conductivity of $\lambda = 0.01 \, \mathrm{W/(K \, m)}$. Right: simulation result for a pebble radius of $R = 0.5 \, \mathrm{mm}$ and a micro-porosity case with a temperature-independent heat conductivity of $\lambda = 0.0018 \, \mathrm{W/(K \, m)}$.}
    \label{fig:daily_variation_both}
\end{figure*}

On circular orbits without obliquity, only a diurnal temperature cycle exists and seasonal effects are absent. For this idealised case with a rotation period of $12.8$ hours (corresponding to the rotation period of 67P, which is also used later), the diurnal surface temperatures at the equator were calculated for two macro-granular cases, i.e. for pebble radii of $R = 5 \, \mathrm{mm}$ and $R=0.5 \, \mathrm{mm}$, respectively, and for a heliocentric distance of $r_H=2  \, \mathrm{AU}$, as shown in Fig. \ref{fig:daily_variation_both}. Several micro-granular cases were also calculated to estimate whether a temperature-independent heat conductivity can also explain the radiative cases. This comparison showed that for pebbles with $R = 0.5  \, \mathrm{mm}$, a constant heat conductivity of $0.0018  \, \mathrm{W/(K \, m)}$ matches the surface temperatures for a diurnal cycle very well (see Fig. \ref{fig:daily_variation_both}, right). The maximum temperature deviation between the macro-porous and micro-porous cases is $\sim 4$~K, briefly after sunrise (see Fig. \ref{fig:daily_variation_both}, right). For larger pebbles with $R = 5  \, \mathrm{mm}$, a reasonable match with a constant heat conductivity of $0.01\, \mathrm{W/(K \,m)}$ can be achieved, but the temperature deviations are more pronounced (see Fig. \ref{fig:daily_variation_both}, left).

\begin{figure*}
    \centering
    \begin{minipage}[b]{.47\linewidth}
    \includegraphics[width=\columnwidth]{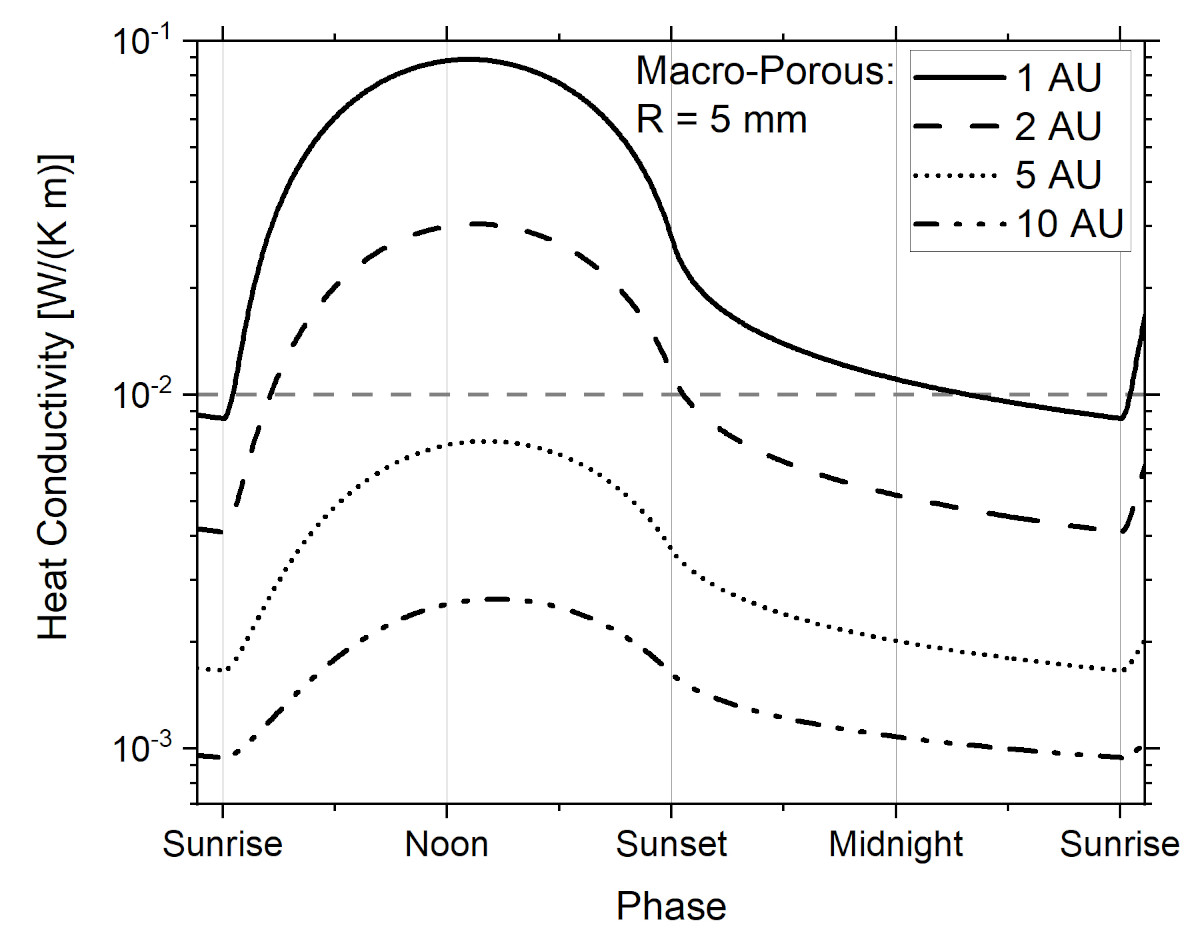}
    \end{minipage}
    \hspace{.05\linewidth}
    \begin{minipage}[b]{.47\linewidth}
    \centering
    \includegraphics[width=\columnwidth]{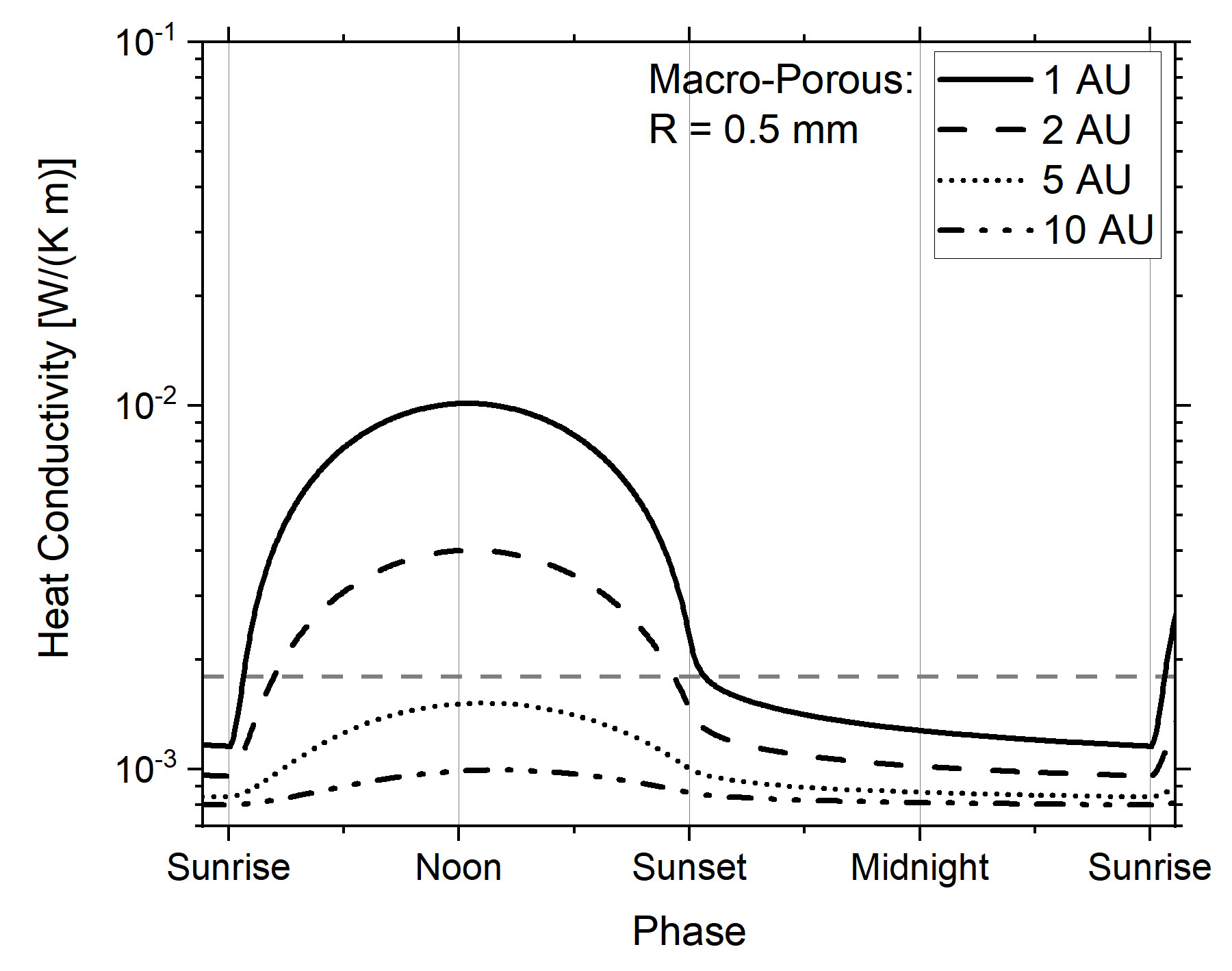}
    \end{minipage}
    \caption{Diurnal cycle of the heat conductivity for different heliocentric distances of $1 \, \mathrm{AU}$ (solid curve), $2 \, \mathrm{AU}$ (dashed curve), $5 \, \mathrm{AU}$ (dotted curve) and $10  \, \mathrm{AU}$ (dashed-dotted curve), respectively. Left: for pebbles with $R = 5  \, \mathrm{mm}$. Right: for pebbles with $R = 0.5 \, \mathrm{mm}$. The horizontal grey dashed lines indicate the micro-granular cases that fit the diurnal surface temperature of the pebble cases for $2  \, \mathrm{AU}$, see Fig. \ref{fig:daily_variation_both}}. 
    \label{fig:daily_variation_both_conductivity} 
\end{figure*}

To illustrate the influence of the heliocentric distance in the macro-porosity cases, Fig. \ref{fig:daily_variation_both_conductivity} shows the diurnal variation of the heat conductivity of the uppermost surface layer for heliocentric distances of $1$, $2$, $5$ and $10 \, \mathrm{AU}$, respectively, for pebble radii of $R = 5\, \mathrm{mm}$ (Fig. \ref{fig:daily_variation_both_conductivity}, left) and $R = 0.5\, \mathrm{mm}$ (Fig. \ref{fig:daily_variation_both_conductivity}, right), respectively. As discussed in Section \ref{sec:thermo_modell}, a smaller pebble radius reduces the radiative heat transport. For larger heliocentric distances, the temperature dependence for the smaller pebbles becomes less important, particularly at nighttime (Fig. \ref{fig:daily_variation_both_conductivity}, right). For the $R = 5\, \mathrm{mm}$ case, radiative heat transfer is still important for large distances and during the full diurnal cycle (see Fig. \ref{fig:daily_variation_both_conductivity}, left). The matching constant values for the micro-granular cases of $r_H=2\, \mathrm{AU}$ (see Fig. \ref{fig:daily_variation_both}) are indicated with the horizontal dashed lines. 

\begin{figure*}
    \centering
    \includegraphics[width=\textwidth]{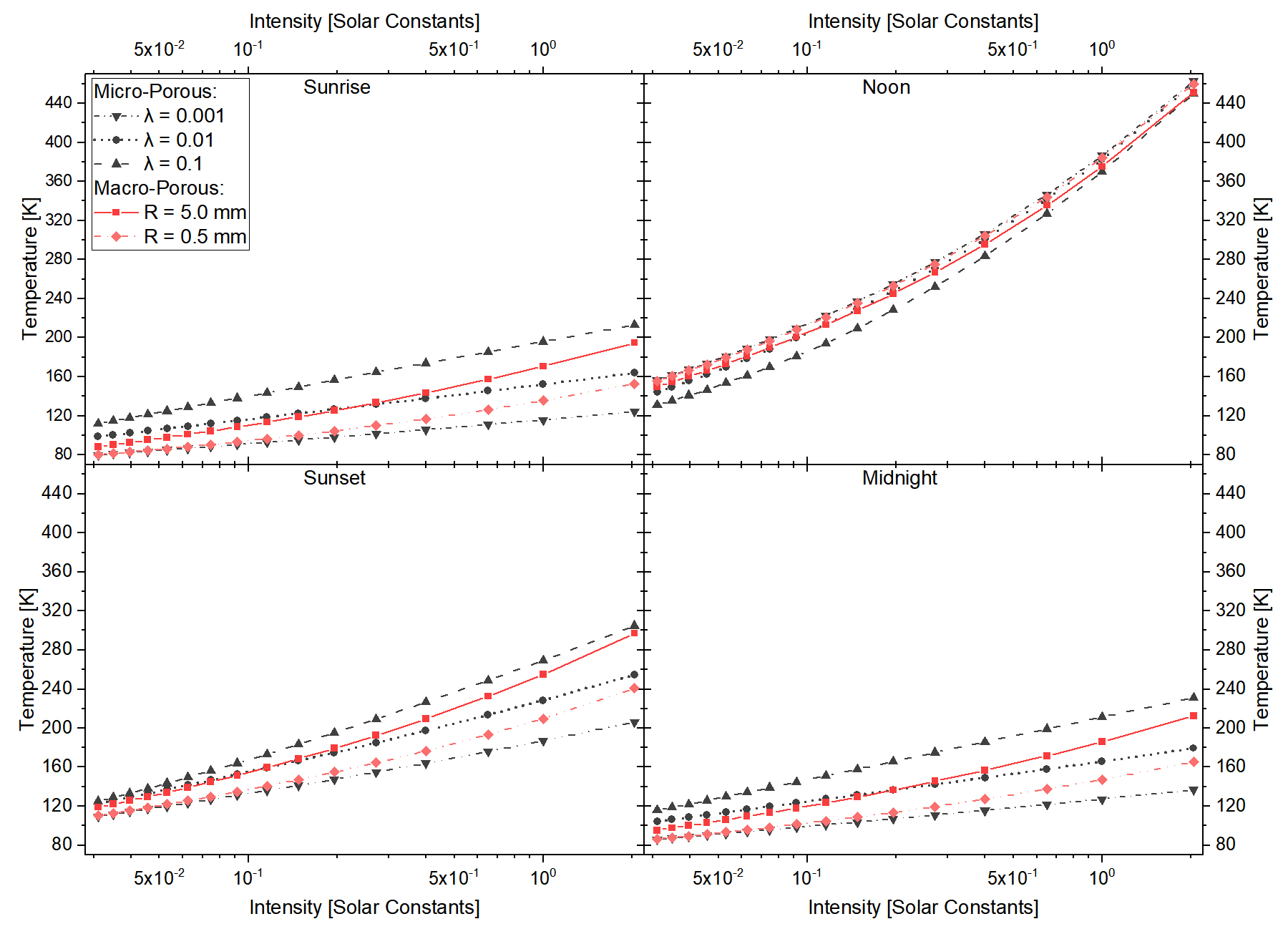}
    \caption{Surface temperature for varying insolation for a circular orbit without obliquity at sunrise (top left), noon (top right), sunset (bottom left) and midnight (bottom right), respectively. The macro-porous pebble cases assume pebble radii of $R = 5\, \mathrm{mm}$ (red squares) and $R = 0.5\, \mathrm{mm}$ (light red diamonds). The micro-porosity cases (dark grey) use a constant heat conductivity of $\lambda_1 = 0.1\, \mathrm{W/(K\, m)}$ (down-pointing triangles), $\lambda_2 = 0.01 \, \mathrm{W/(K\, m)}$ (dots) and $\lambda_3 = 0.001 \, \mathrm{W/(K\, m)}$ (up-pointing triangles), respectively.}
    \label{fig:Circular_Temperature_SolarIntensity}
\end{figure*}

The temperature differences found between both models are very small for most of the diurnal cycle, but they can be larger than the uncertainty of the temperature measurements \citep[e.g. 1.5 K for the MARA sensor;][]{Grott.2019}. However, this difference is not a solid base to distinguish between a constant (micro-granular material) and a strongly temperature-dependent heat conductivity (macro-granular material), because the difference mainly occurs accompanied by strong temperature gradients and it depends on several model parameters, which are mostly not well known. Additionally, features like shadowing, surface roughness and self-heating can also influence the daytime temperatures (see also Sec. \ref{sec:model_validation}), so that mainly nighttime temperatures are usable, but temperature differences at night are too small.
Thus, another measurement strategy had to be found, which is more reliable.
We analysed the surface temperature at the equator resulting from our model at arbitrary diurnal times for varied heliocentric distances, circular orbits and no obliquity. For each heliocentric distance, the two macro-porosity cases (pebble radii $R = 5\, \mathrm{mm}$ and $R = 0.5\, \mathrm{mm}$) and the three micro-porosity cases ($\lambda_1 = 0.1 \, \mathrm{W/(K\, m)}$, $\lambda_2 = 0.01 \, \mathrm{W/(K\, m)}$, and $\lambda_3 = 0.001 \, \mathrm{W/(K\, m)}$) from above were studied. Fig. \ref{fig:Circular_Temperature_SolarIntensity} presents the results for the temperature at sunrise (top left), noon (top right), sunset (bottom left) and midnight (bottom right) as a function of the solar intensity (in solar constants) at noontime. Variation of the noontime insolation is either achieved by considering different heliocentric distances for the circular orbit or by choosing different latitudes for a constant heliocentric distance. Both cases are mathematically identical for our one-dimensional thermophysical model as long as the orbit is circular and the body is spherical and possesses no obliquity. The noontime plot in Fig. \ref{fig:Circular_Temperature_SolarIntensity} reveals that the noon temperatures differ only slightly for four of the five cases simulated. Hence, the noontime temperature is not a suitable diagnostics to distinguish between the models. However, from sunset so sunrise, the temperature differences are relatively large and, more importantly, the results show different dependencies on the noontime insolation. These systematic differences provide the possibility to distinguish between a radiation- and a conduction-dominated heat-transport mechanism. 

\begin{figure}
    \centering
    \includegraphics[width=\columnwidth]{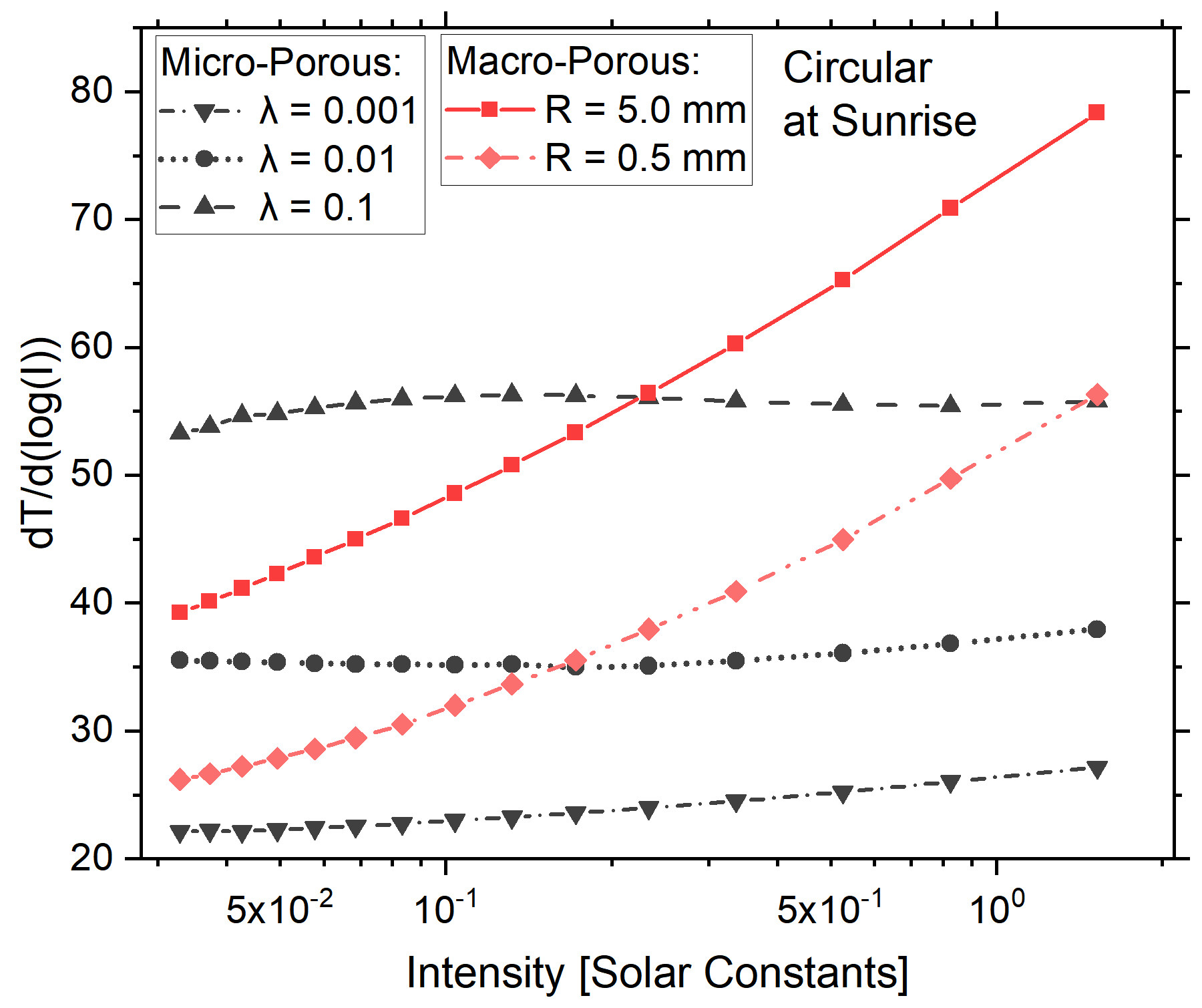}
    \caption{Difference quotient $D$ (Eq. \ref{eq:different_quotient}) as a function of the noontime solar intensity for a circular orbit without obliquity. The macro-granular cases assume pebble radii of $R = 5\, \mathrm{mm}$ (red squares) and $R = 0.5 \, \mathrm{mm}$ (light red diamonds). The micro-granular cases (dark grey) use constant heat conductivities of $\lambda_1 = 0.1 \, \mathrm{W/(K\, m)}$ (down-pointing triangles), $\lambda_2 = 0.01 \, \mathrm{W/(K\, m)}$ (dots) and $\lambda_3 = 0.001 \, \mathrm{W/(K\, m)}$ (up-pointing triangles), respectively.}
    \label{fig:dT_dlogI_circular_sunrise}
\end{figure}

For further analysis, we chose the sunrise temperatures as a proxy for the heat-transport mechanism, because these temperatures show the strongest influence of the heat-conduction process. To mathematically quantify the increase of the surface temperature with varying insolation, we calculated the difference quotient between the sunrise temperatures $T$ and the logarithm of the solar intensity $I$ relative to the solar constant $I_E = 1367 \, \mathrm{W/m^{2}}$,
\begin{equation}
    D = \frac{\Delta T}{\Delta \log (I/I_E)}.
    \label{eq:different_quotient}
\end{equation}
Fig. \ref{fig:dT_dlogI_circular_sunrise} shows the value of $D$ as a function of noontime insolation for our five model cases. For a constant thermal conductivity (micro-granular cases), $D$ remains relatively constant over a wide range of solar intensities. In contrast, the radiation-dominated thermal conductivities (macro-granular cases) result in an intensity dependence of $D$ that becomes stronger with increasing insolation. As this is a major difference in surface-temperature behaviour, it should in principle be measurable by spacecrafts.

\begin{figure*}
    \centering
    \begin{minipage}[b]{.47\linewidth}
    \includegraphics[width=\columnwidth]{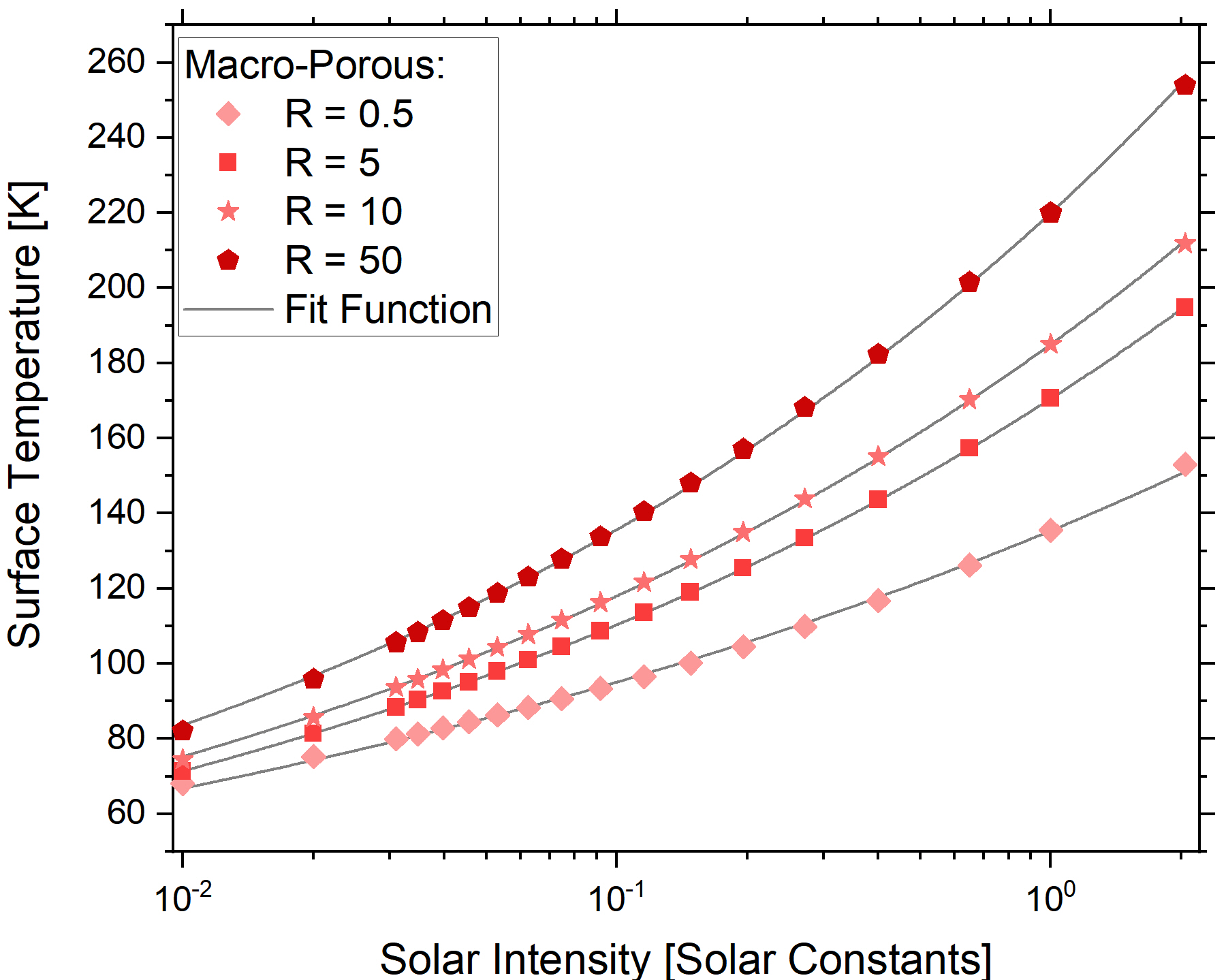}
    \end{minipage}
    \hspace{.05\linewidth}
    \begin{minipage}[b]{.47\linewidth}
    \centering
    \includegraphics[width=\columnwidth]{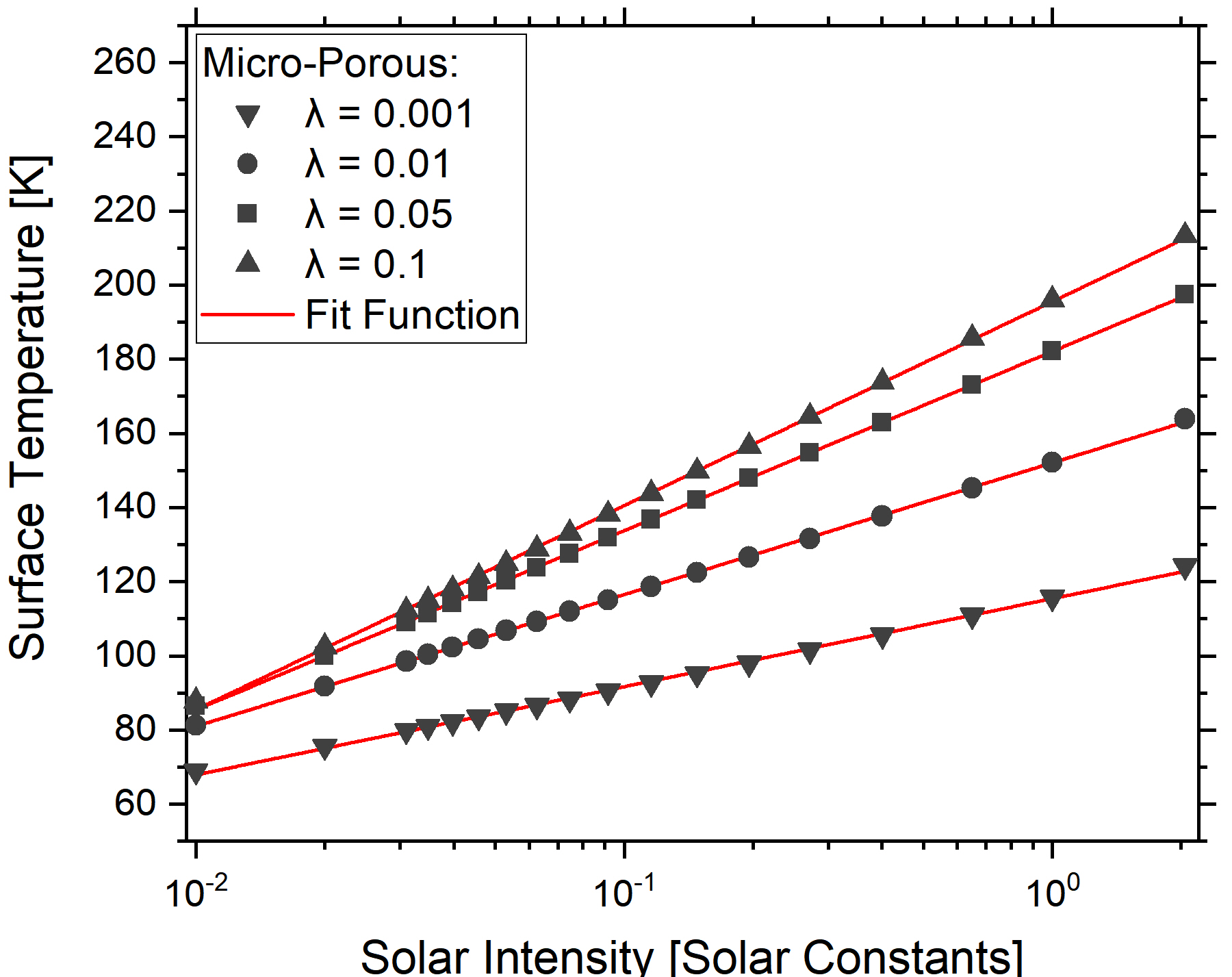}
    \end{minipage}
    \caption{Model results and fit functions for the sunrise temperature as a function of the insolation at noontime. Left: Macro-granular case with pebble radii of $R = 0.5\, \mathrm{mm}$ (diamonds), $R = 5\, \mathrm{mm}$ (squares), $R = 10\, \mathrm{mm}$ (asterisks) and $R = 50\, \mathrm{mm}$ (pentagons) and their corresponding fit function from Eq. \ref{eq:fit_function_radiative}. The corresponding fit parameters $a$ and $b$ can be found in Table \ref{tab:fit_parameter_radiative}. Right: Micro-granular case with heat conductivities $\lambda = 0.1 \, \mathrm{W/(K\, m)}$ (up-pointing triangles), $\lambda = 0.05 \, \mathrm{W/(K\, m)}$ (squares), $\lambda = 0.01 \, \mathrm{W/(K\, m)}$ (dots) and $\lambda = 0.001 \, \mathrm{W/(K\, m)}$  (down-pointing triangles) and their corresponding fit function from Eq. \ref{eq:fit_function_constant}. The corresponding fit parameters $c$ and $d$ can be found in Table \ref{tab:fit_parameter_constant}. }
    \label{fig:Circular_Fitting_Function} 
\end{figure*}

The data shown in Fig. \ref{fig:Circular_Temperature_SolarIntensity} can be fitted by relatively simple analytical functions, as shown in Fig. \ref{fig:Circular_Fitting_Function} for a wider range in pebble radii and constant heat conductivities. For the macro-granular case, we used a power-law function 
\begin{equation}
    T_{\mathrm{sunrise,r}} = a \cdot \left( \frac{I}{I_E} \right)^b.
    \label{eq:fit_function_radiative}
\end{equation}
The fit values of the free parameters $a$ and $b$ are summarised in Table \ref{tab:fit_parameter_radiative} for pebble radii between 0.1~mm and 50~mm. As implied by the constant difference quotient of the micro-granular cases, we applied the following logarithmic function  
\begin{equation}
    T_{\mathrm{sunrise,r}} = c + d \log \left( \frac{I}{I_E} \right) 
    \label{eq:fit_function_constant}
\end{equation}
to fit the data. The fit parameters $c$ and $d$ are presented in Table \ref{tab:fit_parameter_constant} for heat conductivities from $0.001 \, \mathrm{W~m^{-1}~K^{-1}}$ to $0.5 \, \mathrm{W~m^{-1}~K^{-1}}$. In general, for both model cases, the resulting fits are of high quality with a coefficient of determination $R_D^2>0.999$.

\begin{table}
\caption{Fit parameters of the fitting function Eq. \ref{eq:fit_function_radiative} for the macro-granular cases.}
\centering
\begin{tabular}{ccc}
\hline
Pebble Radius & Parameter $a$ & Parameter $b$ \\
{[}mm{]}      & {[}K{]}       & {[}-{]}        \\ \hline
50            &  220.0      & 0.2100  \\
20            &  200.1      & 0.2013  \\
10            &  185.0      & 0.1995  \\
5             &  170.5      & 0.1890  \\
2             &  153.5      & 0.1779  \\
1             &  143.2      & 0.1666  \\
0.8           &  140.4      & 0.1626  \\
0.5           &  135.4      & 0.1536  \\
0.1           &  126.4      & 0.1249  \\ \hline    
\end{tabular}%
\label{tab:fit_parameter_radiative}
\end{table}

\begin{table}
\caption{Fit parameters of the fitting function Eq. \ref{eq:fit_function_constant} for the micro-granular cases. }
\centering
\begin{tabular}{ccc}
\hline
Thermal Cond. & Parameter $c$ & Parameter $d$ \\
{[$\mathrm{W/(K \, m)}$]}      & {[}K{]}       & {[}K{]}        \\ \hline
0.5         & 225.5 & 31.21             \\
0.1         & 195.6 & 23.90             \\
0.05        & 182.1 & 20.99            \\
0.02        & 164.7 & 17.61             \\
0.01        & 152.1 & 15.43             \\
0.006       & 143.2 & 14.04             \\
0.002       & 125.6 & 11.57             \\
0.001       & 115.5 & 10.32         \\ \hline
\end{tabular}%
\label{tab:fit_parameter_constant}
\end{table}

As suggested by the data in Tables \ref{tab:fit_parameter_radiative} and \ref{tab:fit_parameter_constant}, there seems to be a systematic dependency of the fit parameters on the pebble size and thermal conductivity, respectively. These correlations are discussed in Appendix \ref{App:parameter_fitting}. With these correlations, it will be possible to analytically calculate the sunrise temperatures as a function of solar intensity at noontime for any sensible pebble radius or any sensible constant thermal conductivity without the need to run a full set thermophysical simulations. 
\par
\begin{figure*}
    \centering
    \includegraphics[width=\textwidth]{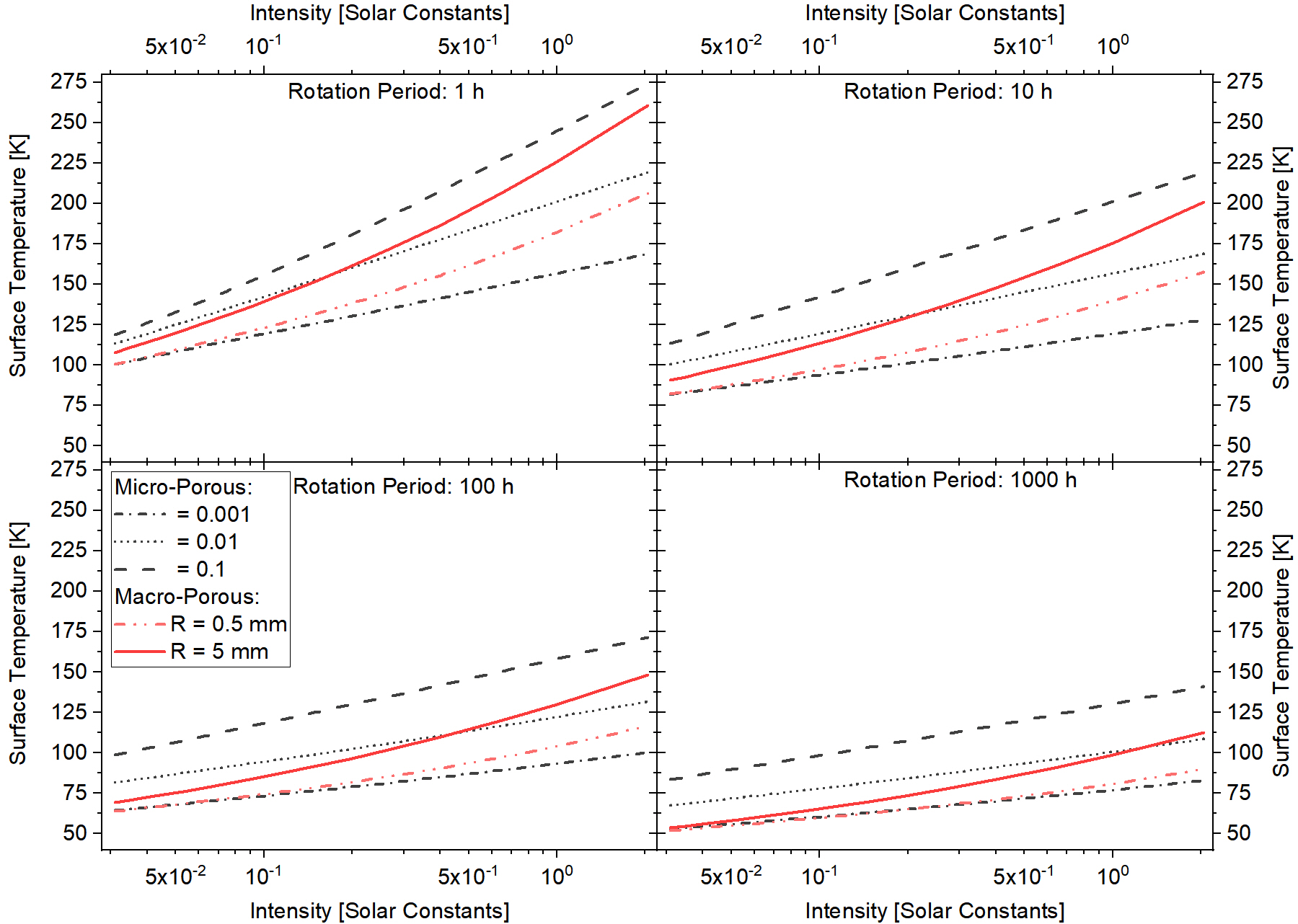}
    \caption{Surface temperatures at sunrise for rotation periods of $1$, $10$, $100$ and $1,000$ hours for varying solar intensities.}
    \label{fig:rotation_variation}
\end{figure*}

Additionally, the rotation period of an object should also have an influence on the sunrise surface temperatures. To investigate this effect, we ran simulations with four different rotational periods, reaching from very fast rotators with $P = 1 \, \mathrm{hour}$ to very slow rotators with $P = 1,000 \, \mathrm{hours}$. The results for the sunrise temperatures of the three micro-porous cases ($\lambda = 0.1, 0.01, 0.001 \, \mathrm{W/(K\, m)}$) and the two macro-porous cases ($R = 0.5, 5\, \mathrm{mm}$) are shown in Fig. \ref{fig:rotation_variation}. Due to runtime constraints, the cases with spin periods of $100$ hours and $1,000$ hours were calculated  for $100$ and $20$ diurnal cycles only, instead of 1,000 cycles. However, the temperature evolution is stable after a few diurnal cycles, as shown in Appendix \ref{App:rotation_variation}. 
\par
The trends in each case are comparable to our previous results, with the tendency that longer spin periods result in smaller sunrise temperatures, reducing the differences between the micro-granular and the macro-granular cases. In general, it can be stated that observations over a larger range of solar intensities (e.g., for more than one order of magnitude) increase the likeliness to differentiate between the two cases by observing the sunrise surface temperatures. In principle, our method could be applied also to the Moon and Mercury, as they are airless bodies covered with regolith. For both bodies, the stratification of the material is influencing the thermophysical processes, as described by \citet[][]{Hayne.2017} for the Moon. However, Moon and Mercury have very long rotation periods, and as we show in Fig. \ref{fig:rotation_variation}, the ability to distinguish between the two cases is reduced with increasing rotation period, so that we did not consider these cases in our work.

\begin{figure}
    \centering
    \includegraphics[width=\columnwidth]{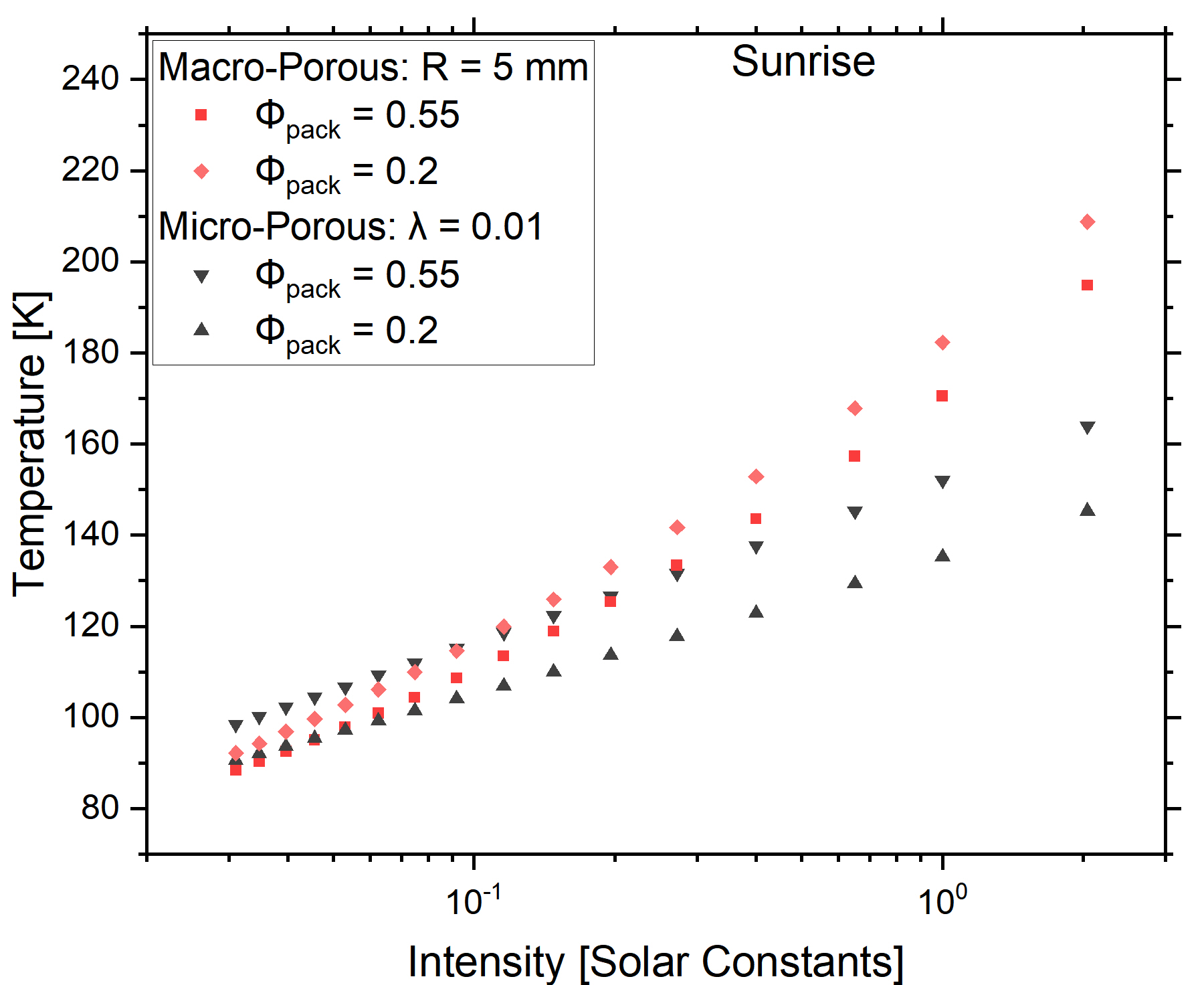}
    \caption{Surface temperatures at sunrise for a volume filling factor of the packing $\Phi_\mathrm{pack} = 0.55$ and $\Phi_\mathrm{pack} = 0.2$ for one macro-granular case (pebble radii of $R = 5\, \mathrm{mm}$) and one micro-granular case (constant heat conductivities of $\lambda = 0.01 \, \mathrm{W/(K\, m)}$), respectively.}
    \label{fig:comparison_porosity}
\end{figure}

As described in Sec. \ref{sec:thermo_modell}, we used a default volume filling factor of the packing of $\Phi_\mathrm{pack} = 0.55$. To address the stratification of the material, we also ran model I with a constant thermal conductivity of $\lambda = 0.01 \, \mathrm{W/(K\, m)}$ and model II with a pebble radius of $R = 5 mm$ with a volume filling factor $\Phi_\mathrm{pack} = 0.2$, which could be the case for a less dense surface. This variation does not directly include a depth-dependency of the material properties, but it corresponds to the two extremes of a very porous structure and a densely packed material. The bulk density of the body is reduced from $\rho = 532 \, \mathrm{Kg/m^3}$ to $\rho = 193 \, \mathrm{Kg/m^3}$ for $\Phi_\mathrm{pack} = 0.2$. The results are plotted in Fig. \ref{fig:comparison_porosity}. For higher porosities, the mean-free path in the macro-porous case is increased, resulting in an even higher radiative heat conductivity. Therefore, the sunrise temperatures are higher than in the denser case and increase with a steeper slope. In contrast, the temperatures in the micro-porous case are reduced due to the change in density. The dependency on the solar intensity only changes slightly. Hence, we assume that if a surface shows a stratification, the dependency of the sunrise temperature on the noon intensity still offers the opportunity to distinguish the micro-granular from the macro-granular structure.

\subsection{A Realistic Case}
\label{sec:results_sub:realistic}

For a more realistic application of our model, we chose to use the orbital parameters of comet 67P. The start of a simulation run is set to the aphelion position of comet 67P at $5.68$ AU, whereupon we follow the comet for $1.5$ orbital periods, but only data from perihelion to perihelion is used for the analysis. Hence, the first half of the orbit is used for thermal equilibration of the model. 
\par 
Analogue to the analysis in Section \ref{sec:results_sub:ideal}, the surface temperature is calculated for varying illumination intensities. These include the decrease or increase due to the variation of the heliocentric distance and the migration of the sub-solar point, due to comet 67P's obliquity of $52^\circ$. The observed point was set to the equator, hence the migration of the sub-solar point does not include a variation of the length of insolation per diurnal cycle. As shown before, the rotation period or length of the sunlit part of the diurnal cycle has an influence on the absolute temperatures, which would be also expected in the oblique case with an observation point elsewhere than on the equator.   
\par
\begin{figure}
    \centering
    \includegraphics[width=\columnwidth]{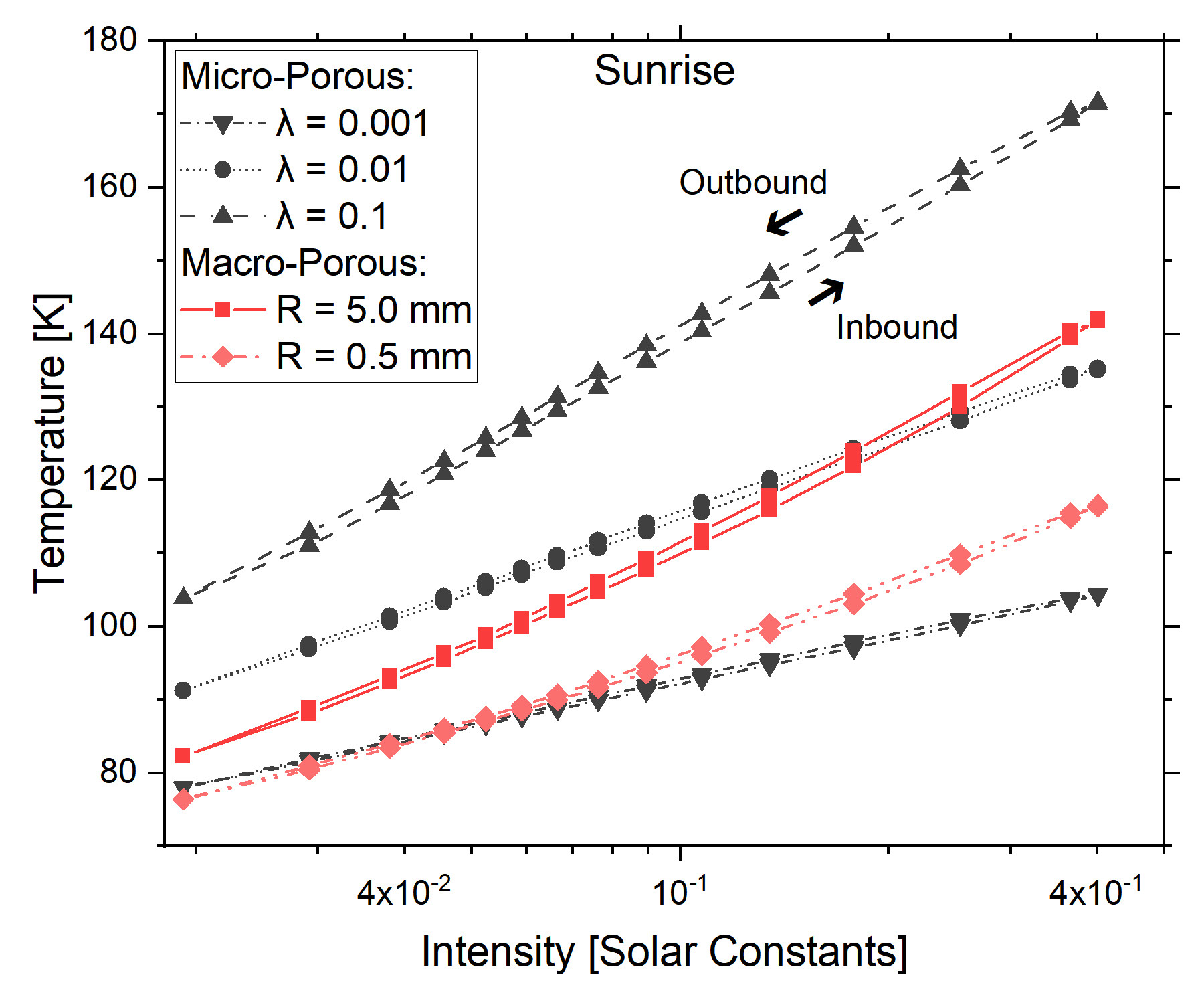}
    \caption{Sunrise surface temperatures for varying insolation for the orbit of comet 67P with an obliquity of $52^\circ$. The macro-granular pebble cases assume pebble radii of $R = 5 \, \mathrm{mm}$ (red squares) and $R = 0.5\, \mathrm{mm}$ (light red diamonds). The micro-granular cases (dark grey) use constant heat conductivities of $\lambda = 0.1 \, \mathrm{W/(K\, m)}$ (down-pointing triangles), $\lambda = 0.01 \, \mathrm{W/(K\, m)}$ (dots) and $\lambda = 0.001 \, \mathrm{W/(K\, m)}$ (up-pointing triangles).}
    \label{fig:Axial_tilt_sunrise_Temperature_SolarIntensity}
\end{figure}

As discussed in Section \ref{sec:results_sub:ideal}, the surface temperatures at sunrise are most suitable to study the difference of heat conduction mechanisms. Therefore, we plotted these temperatures in Fig. \ref{fig:Axial_tilt_sunrise_Temperature_SolarIntensity}, whereas the surface temperatures at noon, sunset and midnight can be found in Appendix \ref{App:orbital_scenarios} (Fig.  \ref{fig:Axial_Tilt_Temperature_SolarIntensity_Elliptical}) for the sake of completeness. 

\begin{figure}
    \centering
    \includegraphics[width=\columnwidth]{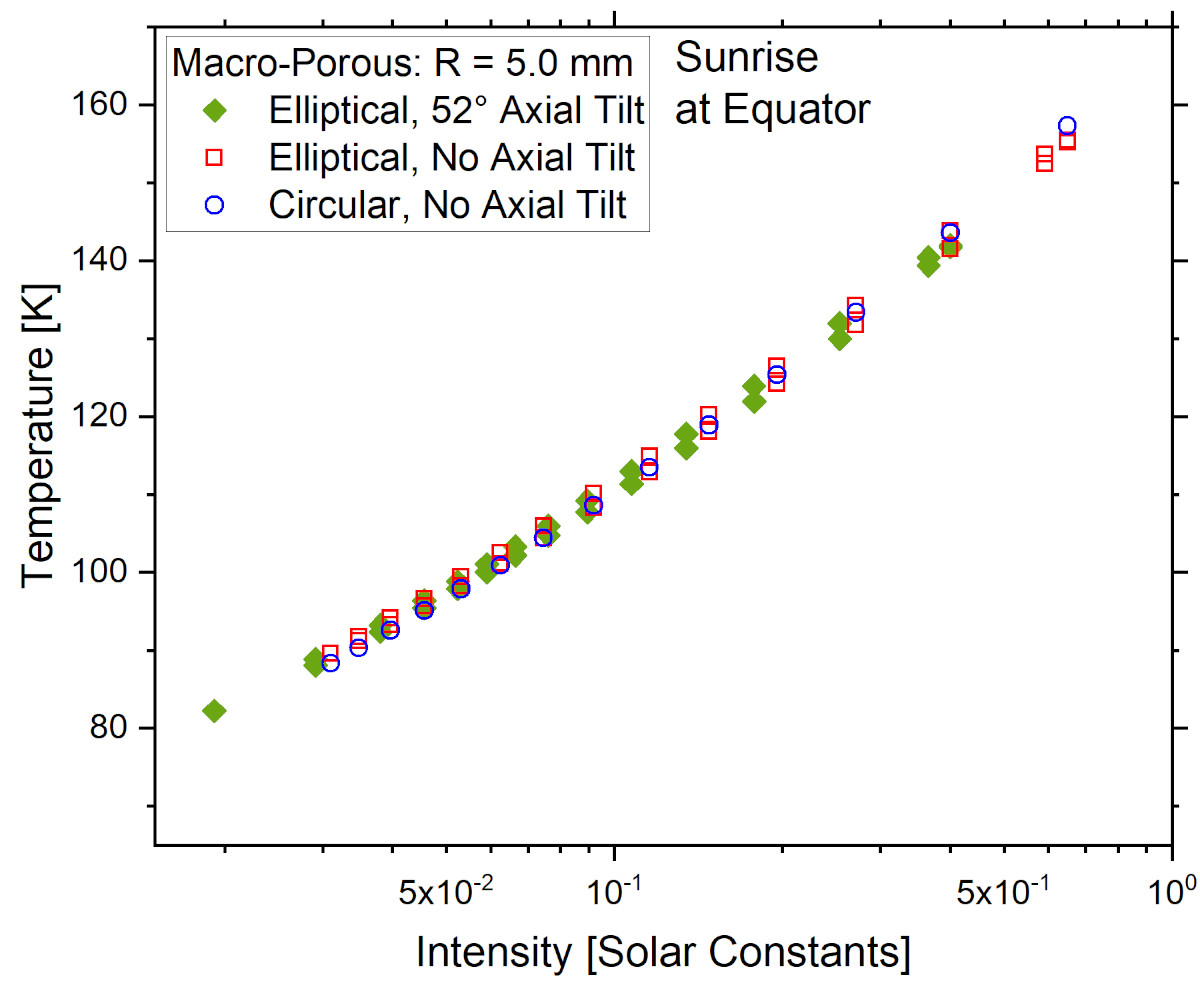}
    \caption{Sunrise surface temperatures for varying insolation for the macro-granular case with $R = 5 \, \mathrm{mm}$ pebble radius and three different orbital scenarios, i.e. a circular orbit (blue open circles), an elliptical orbit without obliquity (red open squares) and an elliptical orbit with an obliquity of $52^\circ$ (green diamonds).}
    \label{fig:SunriseTemperature_SolarIntensity_orbital_scenarios}
\end{figure}

In contrast to the circular orbit, the difference of approaching the Sun before perihelion and departing from the Sun afterwards is clearly visible in the data as a hysteresis. This slight asymmetry is caused by stored energy in the system after perihelion passage. However, the same general dependency as in the circular scenario for the two model cases can still be observed, with a much steeper dependency of the sunrise temperature on the insolation at noontime for the macro-porosity case. This suggests that the exact orbital setup of a body has only a minor influence on the surface temperature cycles. This effect is shown in Fig. \ref{fig:SunriseTemperature_SolarIntensity_orbital_scenarios} where all temperature data for a specific macro-granular case, i.e. a specific pebble radius, collapse on one curve. This also shows that the influence of orbital eccentricity and obliquity on the sunrise temperature for constant noontime insolation is small. 

\par 
Analogue to Fig. \ref{fig:dT_dlogI_circular_sunrise} for the circular scenario, a constant difference quotient $D$ (Eq. \ref{eq:different_quotient}) is present in the micro-granular cases for elliptical orbits (compare Fig. \ref{fig:Axial_tilt_sunrise_Temperature_SolarIntensity} with Fig. \ref{fig:Circular_Temperature_SolarIntensity}). This underlines that observing the surface temperature at sunrise for varying illumination conditions at noontime is a suitable strategy to investigate the surface structure of small Solar System objects by remote measurements. 

\section{Summary and Conclusions}
\label{sec:conclusions}
In Section \ref{sec:results_sub:ideal}, we showed that studying the surface temperature of a single point on the surface over a full diurnal cycle cannot be used to infer the surface granularity of the material. In case of a macro-granular surface material, for which the radiative (i.e. temperature-dependent) energy transport plays an important role, the diurnal temperature curve can still be fitted by assuming a constant (i.e. temperature-independent) heat conductivity (see Figure \ref{fig:daily_variation_both}). 
\par
However, we also showed that it is possible to distinguish between two extreme types of granularity in planetary surfaces, namely between a micro-granular regolith, for which the dominating heat-transfer process is inter-grain conduction, which is only moderately temperature-dependent, and a macro-granular regolith, whose pore space is large enough to allow the strongly temperature-dependent radiative energy transport to dominate over conduction, at least during daytime. We demonstrated that this distinction can be measured through the determination of the sunrise temperature as a function of the insolation at noontime. The correlation between these two properties is characteristic for the heat-transport mechanism, leading to the dependencies of the sunrise temperature on the noontime insolation shown in Eqs. \ref{eq:fit_function_radiative} and \ref{eq:fit_function_constant}. In principle, the parameters $a$ and $b$ in Eq. \ref{eq:fit_function_radiative} can be used to determine the pebble size (see Table \ref{tab:fit_parameter_radiative}), when the rotation period is known (see Appendix \ref{App:rotation_variation}). Similarly, the parameters $c$ and $d$ in Eq. \ref{eq:fit_function_constant} may be utilised for the determination of the heat conductivity of a micro-porous regolith (see Table \ref{tab:fit_parameter_constant} and Appendix \ref{App:rotation_variation}).
\par
We showed in Fig. \ref{fig:Axial_tilt_sunrise_Temperature_SolarIntensity} that the cause for the variable insolation at noontime has a negligible influence on the sunrise temperature, so that either different heliocentric distances for eccentric orbits (e.g. for cometary nuclei) or latitudinal or seasonal variations may be used. Hence, our method can be utilised by escorting space missions as well as by flyby missions, like ESA's Comet Interceptor  \citep{Snodgrass.2019}, as long as sufficient latitudinal coverage for the measurements of the sunrise temperature can be ensured.  
\par
Unfortunately, the available sunrise surface temperature data of small Solar System bodies is limited. As explained in Sec. \ref{sec:introduction}, VIRTIS data cannot be used, due to its lower limit of measurement capability. The comparison to MIRO data is almost impossible, because of different aspects. First, due to its long wavelengths, the MIRO instrument measures heat fluxes from deeper and shallower layers simultaneously. Thus, a comparison would require the usage of a radiative transfer model to retrieve the real surface temperature, which is beyond the scope of this work. Second, the data must be restricted to low emission angles and low distances between comet and space craft. Applying this condition, only data for high heliocentric distances remain and the small variation of the solar intensity does not allow to differentiate between the two cases. Our excursion to the MIRO measurements has shown that finding suitable data is complicated, because high solar intensities and high spatial resolution are difficult to achieve simultaneously. This is why we propose that future space missions should follow this measurement strategy to allow for a remote analysis of the surface structure of small Solar System objects. Additionally, in case of asteroids and the surprising observation of large boulders on Ryugu and Bennu, our method could potentially be used to distinguish between those boulders and the regolith-covered surface, depending on the dominating heat transport mechanism. 
\par
However, there are also limitations of our proposed method. For large heliocentric distances and small pebble sizes, a distinction between the insolation dependencies of the sunrise temperatures, as shown in Eqs. \ref{eq:fit_function_radiative} and \ref{eq:fit_function_constant}, becomes increasingly difficult, due to the shallowness of the power-law function in Eq. \ref{eq:fit_function_radiative}. This might limit the potential applicability of our method to bodies not further away from the Sun than the asteroid belt.

\section{Future Work}
\label{sec:outlook}
We adopted a number of simplifications in this work, which might have implications on the result and need to be taken into account in future modelling:
\begin{itemize}
    \item It needs to be noted that the influence of surface roughness as well as shadowing and self-heating effects have been neglected in the present work. This might be an oversimplification, especially at daytimes, as indicated for example by \citet[][]{Grott.2019}.
    \item Our model neglects the sublimation of volatiles, which is valid for several small bodies. For comets, volatiles play an important role and water ice was observed at the surface of 67P \citep{Filacchione.2016b}. However, most of the surface is ice-free and our model could be applicable to these regions. To understand the influence of ices on our method, further work is need.
    \item We assumed that the heat conductivity in the micro-granular case is temperature-independent, which is a valid assumption since the temperature dependency only stems from the material parameters, namely from the thermal conductivity of the solid material, which is generally not a strong effect \citep[a factor of $\sim 2$ per $200\, \mathrm{K}$, see Fig. 3 in][]{Opeil.2020}. However, even for a slight temperature dependency, as long as it is weaker than that of the macro-porous case, our proposed method should still be feasible, if the observed solar intensity variation is large enough. Future work on this is certainly welcome.
    \item We also assumed that the regolith properties do not change with depth. Especially small bodies ($< 100 \, \mathrm{km}$ in size), like comets and small asteroids, should not have experienced any strong material stratification. In case of comets, radar measurements indicate a less porous structure at the subsurface on a scale of ten metres \citep[][]{Kofman.2020}. However, for the uppermost centimetres of the material, which are important in our model, no constraint is given for depth-depending material properties by direct measurements. Due to the ongoing activity and surface erosion of comets, less altered material will be exposed at the surface. Larger objects like large asteroids or the largest comets might have experienced material compaction. This is also true for airless bodies, like the Moon and Mercury. Density and volume filling factor increase with increasing depth, influencing the thermal properties. For the Moon, the thermal conductivity varies mostly in the first ten centimetres. At deeper layers, it is assumed to approach a constant value. The increase of the thermal conductivity because of a density increase, from the surface to a depth of $10 \,\mathrm{cm}$, is roughly a factor of $4-5$ \citep[][]{Hayne.2017}. To address a possible stratification, we used two extreme cases, a densely packed structure and a more porous one. For both cases, our proposed model is applicable and hence we assume that this is also true for the transition between both. However, to verify this assumption it should be investigated how depth-dependent properties change the surface temperature. A good start in this direction would be the regolith-stratification work by \citet{Schrapler.2015}.
\end{itemize}

\section*{Acknowledgements}

This work was funded through the DFG project BL 298/27-1. 

\section*{Data Availability}
The data underlying this article will be shared on reasonable request to the corresponding author.




\bibliographystyle{mnras}
\bibliography{paper_references} 



\appendix

\section{Details Of The Thermophysical Model}
\label{App:thermophysical_model}
As described in Section \ref{sec:thermo_modell}, we used a thermophysical model that comprises of heat transport by conduction through the grain network (micro- and macro-porosity cases) as well as by radiation through the void spaces (macro-porosity case only). The radiative heat conductivity can be written as
\begin{equation}
    \lambda_\mathrm{rad}(T) = \frac{16}{3} \sigma \, l \, T^3 ,
\end{equation}
with the Stefan-Boltzmann constant $\sigma$ and the mean free path $l$ inside the voids. We assume that the mean free path depends on the pebble radius $R$ and on the volume filling factor of the pebble packing $\Phi_\mathrm{pack}$, 
\begin{equation}
    l = e R \frac{1-\Phi_\mathrm{pack}}{\Phi_\mathrm{pack}} \, \mathrm{,}
\end{equation}
here the scaling factor measures $e=1.34$, which was empirically estimated \citep[see][]{Gundlach.2012}. Radiative energy transport inside the pebbles is neglected, due to the minuscule void spaces inside the pebbles. 
\par
The network conductivity describes the conduction through the material. In case of a granular material, the Hertz factor describes the reduced contact area between grains. Here, the word grains describes the small particles building the micro-porous material or the pebbles. The network thermal conductivity reads
\begin{equation}
    \lambda_\mathrm{net, micro} = \lambda_\mathrm{par} \, H_\mathrm{micro},
\end{equation}
with the material thermal conductivity, $\lambda_\mathrm{par}$, and the Hertz factor, $H_\mathrm{micro}$, which is given by
\begin{equation}
    H_\mathrm{micro} = \left[\frac{9 (1 - \mu_\mathrm{par}^2)}{4 E_\mathrm{par}} \pi \gamma_\mathrm{par} r^2 \right]^{1/3} \xi(\Phi_\mathrm{micro},r).
\end{equation}
Here, $\mu_\mathrm{par}$, $E_\mathrm{par}$ and $\gamma_\mathrm{par}$ are the Poisson ratio, the Young's modulus and the specific surface energy of a particle, respectively, $r$ denotes the dust grain radius and $\Phi_\mathrm{micro}$ the volume-filling factor of the micro-porous structure. $\xi(\Phi_\mathrm{micro},r)$ is an empirical factor, that describes the packing geometry of the particles, which depends on the volume-filling factor of the micro-porous structure and on the particle size \citep[][]{Gundlach.2012}, 
\begin{equation}
    \xi(\phi_{\rm micro},r) = \frac{f_1 \, \exp\left[f_2 \, \phi_{\mathrm{micro}}\right]}{r} \, .
\end{equation}
Here, the two parameters are $f_1 = 5.18 \cdot 10^{-2}$ and $f_2 = 5.26$.
\par
For macro-porous structures built up by the pebbles, the network conductivity is significantly reduced, due to the reduced contact area between the pebbles. Therefore, a second Hertz factor $H_\mathrm{macro}$ is introduced \citep[][]{Gundlach.2012}, whose calculation is based on the parameters of the pebble structure,
\begin{equation}
    H_\mathrm{macro} = \left[\frac{9 (1 - \mu_\mathrm{agg}^2)}{4 E_\mathrm{agg}} \pi \gamma_\mathrm{agg} R^2 \right]^{1/3} \xi(\Phi_\mathrm{pack},R).
\end{equation}
Here, $\mu_\mathrm{agg}$ is the Poisson ratio, $E_\mathrm{agg}$ the Young's modulus and $\gamma_\mathrm{agg}$ the specific surface energy of a pebble, respectively. The parameter $\xi(\Phi_\mathrm{pack},r)$ is defined analogously as mentioned above. The specific surface energy of a pebble, $\gamma_\mathrm{agg}$, deviates from that of a single particles, due to the porous structure. Referring to \citet{Gundlach.2012}, it reads
\begin{equation}
    \gamma_\mathrm{agg} = \Phi_\mathrm{micro} \, \gamma_\mathrm{par}^{5/3} \left[\frac{9 \pi (1 - \mu_\mathrm{agg}^2)}{r E_\mathrm{par}} \right]^{2/3}.
\end{equation}
With these factors derived above, we can describe the network conductivity in the pebble case as
\begin{equation}
    \lambda_\mathrm{net, macro} = \lambda_\mathrm{par} \, H_\mathrm{micro} \, H_\mathrm{macro}.
\end{equation}
The simulations are performed by implementing these equations into our code. The used material parameters and physical properties are summarized in \ref{Tab:1_Parameters}. Some parameters, e.g. the specific surface energy, the Poisson ratio or the Young’s modulus are temperature dependent. Due to simplicity and with the assumption that these dependencies are not crucial, we neglect them here. Therefore, for a fixed pebble radius, the network conductivity remains constant over time and depth in a simulation run.

\begin{table*}
\caption{Physical parameters used in this study. To be able to compare the simulation results with Rosetta data, properties of comet 67P were chosen.}
\begin{tabular}{lcccc}
\hline
Parameter & Symbol & Value & Unit & Reference \\ \hline
Solar constant & $I_E$ & $1367 $ & $\mathrm{W \, m^{-2}}$ & - \\
Stefan–Boltzmann constant & $\sigma$ & $ 5.67 \times 10^{-8} $&$ \mathrm{W \, m^{-2} \, K^{-4}} $ & -\\
Boltzmann constant & $k$ & $1.38 \times 10^{-23} $&$ \mathrm{J \, K^{-1}}$ & - \\
Bulk density & $\rho$ & $532 $&$ \mathrm{kg \, m^{-3}}$ & \citet{Jorda.2016} \\
Heat capacity & $c_0$ & $560 $&$ \mathrm{J \, kg^{-1} K^{-1}}$  & \citet[][]{Waples.2004} \\
Albedo & $A$ & $0.055$ & - & \citet{Sierks.2015} \\
Emissivity & $\epsilon$ & 1 &  - & - \\
Dust particle radius & $r$ & $0.1 $&$ \mathrm{\mu m}$ &    \citet{Mannel.2016,Mannel.2019}    \\
Pebble radius & $R$ &  - & $\mathrm{mm}$ & Free parameter     \\
Volume filling factor of pebble packing & $\Phi_\mathrm{pack}$ & $0.55$ & - & \citet{Blum.2014,ORourke.2020}     \\
Inter-pebble volume filling factor & $\Phi_\mathrm{agg}$ & $0.4$  & - & \citet{Weidling.2009}   \\
Poisson ratio of pebble & $\mu_\mathrm{agg}$  & $0.17$ & -   & \citet{Weidling.2012}   \\
Poisson ratio of particle & $\mu_\mathrm{par}$  & $0.17$ & -  & \citet{Chan.1973}    \\
Young's modulus of pebble & $E_\mathrm{agg}$  & $8.1 \times 10^{3} $&$ \mathrm{Pa}$   & \citet{Weidling.2012}   \\
Young's modulus of particle & $E_\mathrm{par}$  & $5.5 \times 10^{10} $&$ \mathrm{Pa}$ & \citet{Chan.1973}      \\
Specific surface energy of particle & $\gamma_\mathrm{par}$  & $0.1 $&$ \mathrm{J \, m^{-2}}$ & \citet{Heim.1999}      \\
Heat conductivity of particle & $\lambda_\mathrm{par}$ & $0.5 $&$ \mathrm{W \, m^{-1} K^{-1}}$ & \citet{Blum.2017} \\
Packing structure coefficient & $f_1$ & $5.18 \times 10^2$ & -  & \citet{Gundlach.2012} \\
 & $f_2$ & $5.26$ & -  & \citet{Gundlach.2012} \\
Mean free path coefficient & $e$  & $1.34$ & -  & \citet{Gundlach.2012} \\
\hline
\end{tabular}
\label{Tab:1_Parameters}
\end{table*}

\section{Fit Functions For The Ideal Case}
\label{App:parameter_fitting}
As described in Section \ref{sec:results_sub:ideal}, we fitted functions to each model result (see Eq. \ref{eq:fit_function_radiative} and \ref{eq:fit_function_constant}). From these fit functions, we can investigate the dependency of the free parameters of the fit functions on the model variables, namely the pebble radius (macro-porosity case) and the constant thermal conductivity (micro-porosity case). The results are shown in Figs. \ref{fig:Parameter_A_Pebble} and \ref{fig:Parameter_B_Pebble} for the parameters $a$ and $b$ for the macro-porosity case, and in Figs. \ref{fig:Parameter_C_const} and \ref{fig:Parameter_D_const} for the parameters $c$ and $d$ in the micro-porosity case.

\subsection{Macro-Porosity Case}
\begin{figure}
    \centering
    \includegraphics[width=\columnwidth]{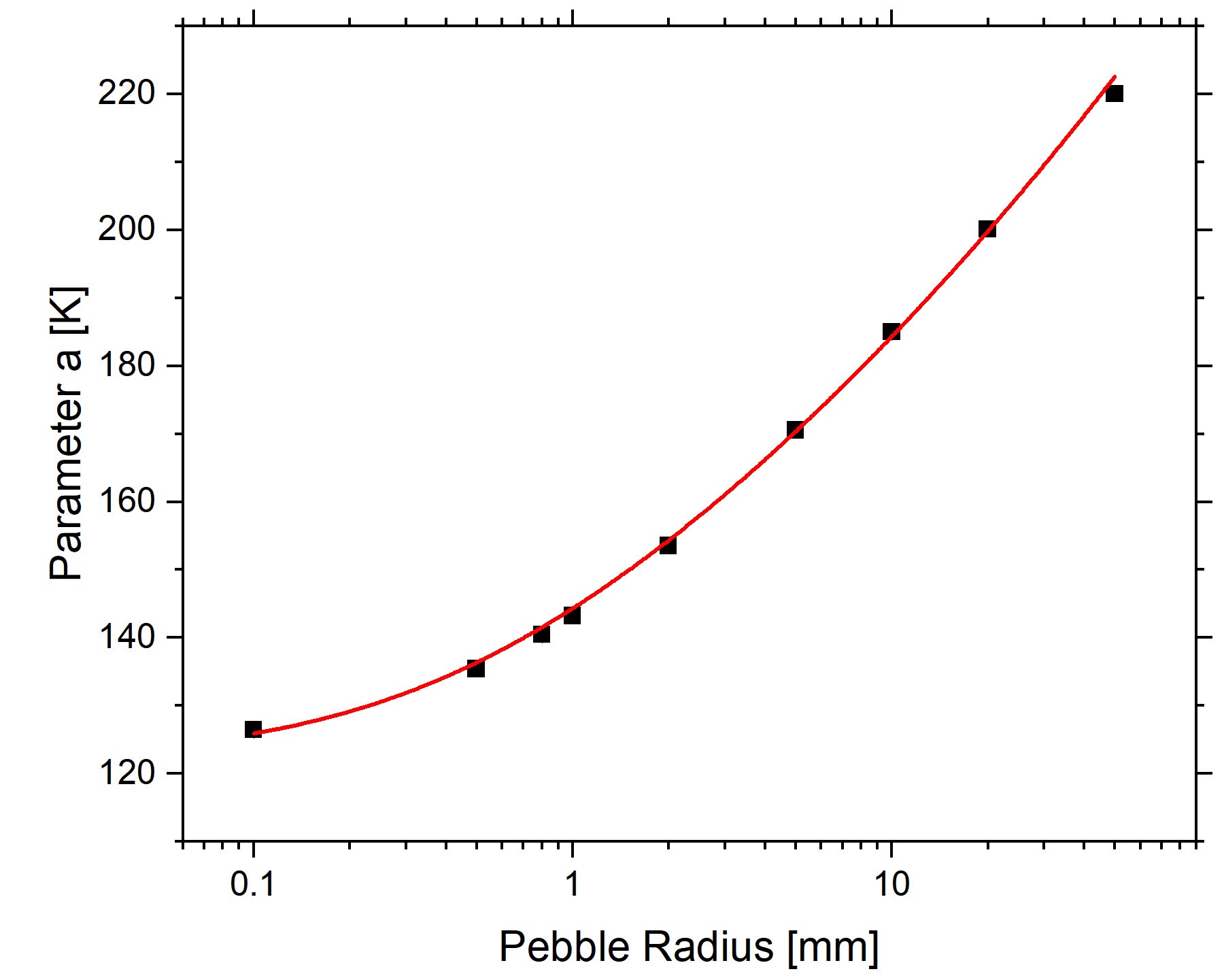}
    \caption{Parameter $a$ (symbols) from Eq. \ref{eq:fit_function_radiative} in the macro-porosity case for different pebble radii $R$. Their relation is approximated by Eq. \ref{eq:parameter_a_pebble} (solid curve). The goodness of the fit is given by a coefficient of determination of $R_D^2 = 0.998$.}
    \label{fig:Parameter_A_Pebble}
\end{figure}

\begin{figure}
    \centering
    \includegraphics[width=\columnwidth]{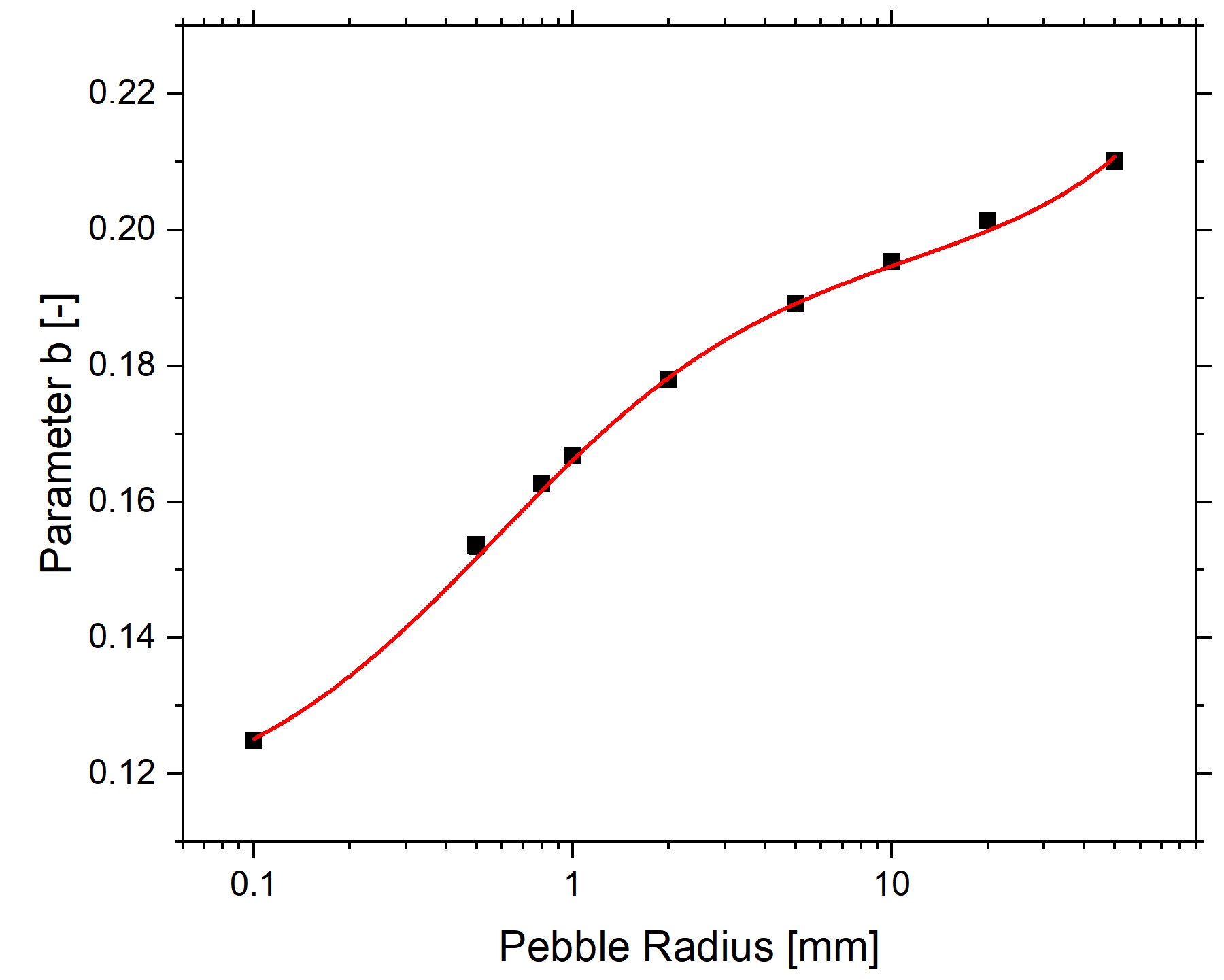}
    \caption{Parameter $b$ (symbols) from Eq. \ref{eq:fit_function_radiative} in the macro-porosity case for different pebble radii $R$. Their relation is approximated by Eq. \ref{eq:parameter_b_pebble} (solid curve). The goodness of the fit is given by a coefficient of determination of $R_D^2 = 0.999$.}
    \label{fig:Parameter_B_Pebble}
\end{figure}

As can be seen in Figs. \ref{fig:Parameter_A_Pebble} and \ref{fig:Parameter_B_Pebble}, there is a systematic dependency of the model parameters $a$ and $b$ from Eq. \ref{eq:fit_function_radiative} on the pebble radius. In order to allow an extrapolation of our model to arbitrary pebble radii, we fitted the empirical data with two simple analytic functions of the form
\begin{equation}
    a = 139.66 \, \mathrm{K} \, \left(\frac{R}{1 \, \mathrm{mm}}  + 0.32 \right)^{0.119}
    \label{eq:parameter_a_pebble}
\end{equation}
and
\begin{equation}
    b = \frac{0.112 + 0.342 \frac{R}{1 \, \mathrm{mm}}}{1 + 1.742 \frac{R}{1 \, \mathrm{mm}} - 0.0025 \left( \frac{R}{1 \, \mathrm{mm}} \right)^2}.
    \label{eq:parameter_b_pebble}
\end{equation}
The solid red lines in Figs. \ref{fig:Parameter_A_Pebble} and \ref{fig:Parameter_B_Pebble} show the quality of the two fit functions. 


\subsection{Micro-Porosity Case}
\begin{figure}
    \centering
    \includegraphics[width=\columnwidth]{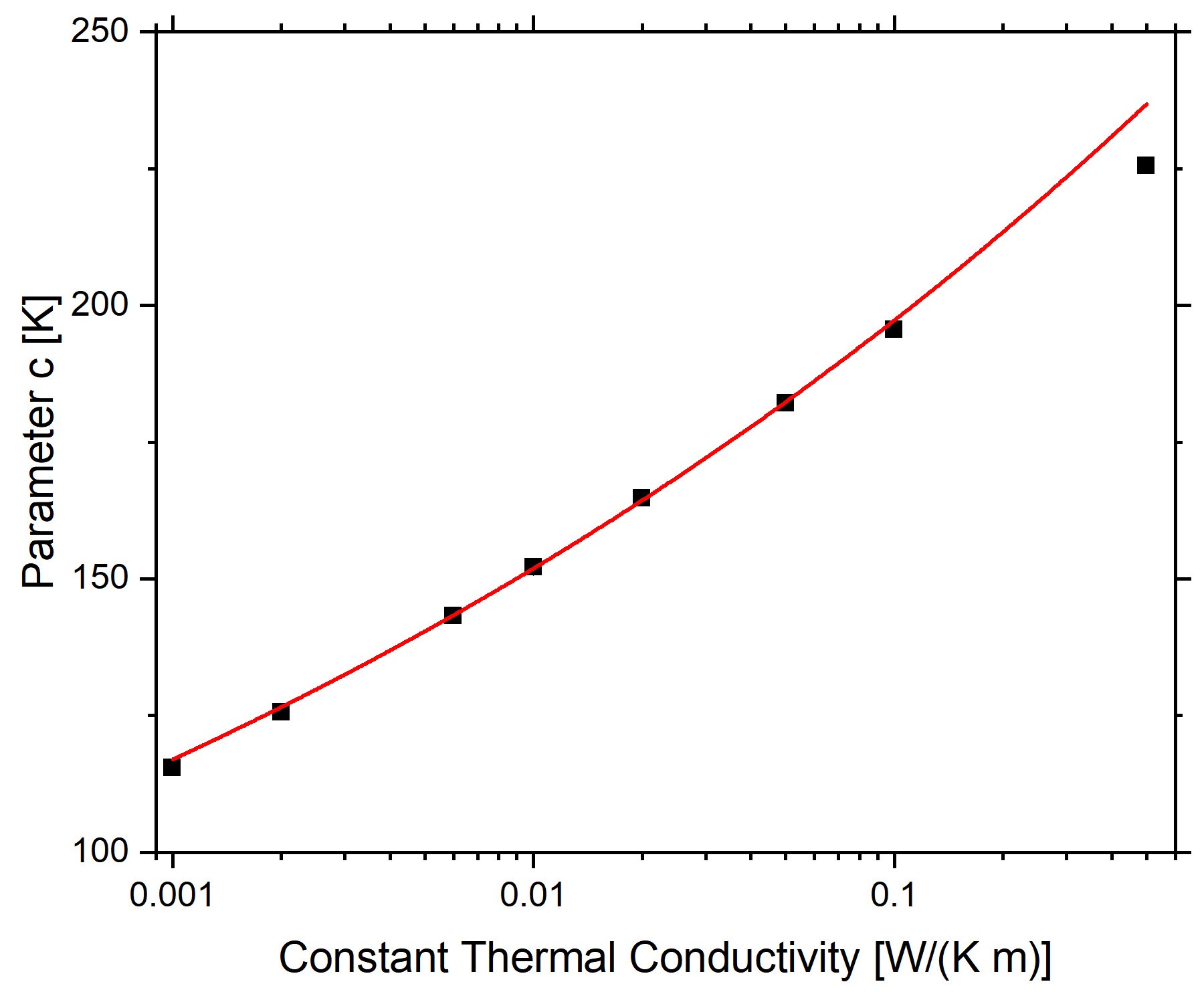}
    \caption{Parameter $c$ (symbols) from Eq. \ref{eq:fit_function_constant} in the micro-porosity case for different constant thermal conductivities $\lambda$. Their relation is approximated by Eq. \ref{eq:parameter_c_const} (solid curve). The goodness of the fit is given by a coefficient of determination of $R_D^2 = 0.998$. } 
    \label{fig:Parameter_C_const}
\end{figure}

\begin{figure}
    \centering
    \includegraphics[width=\columnwidth]{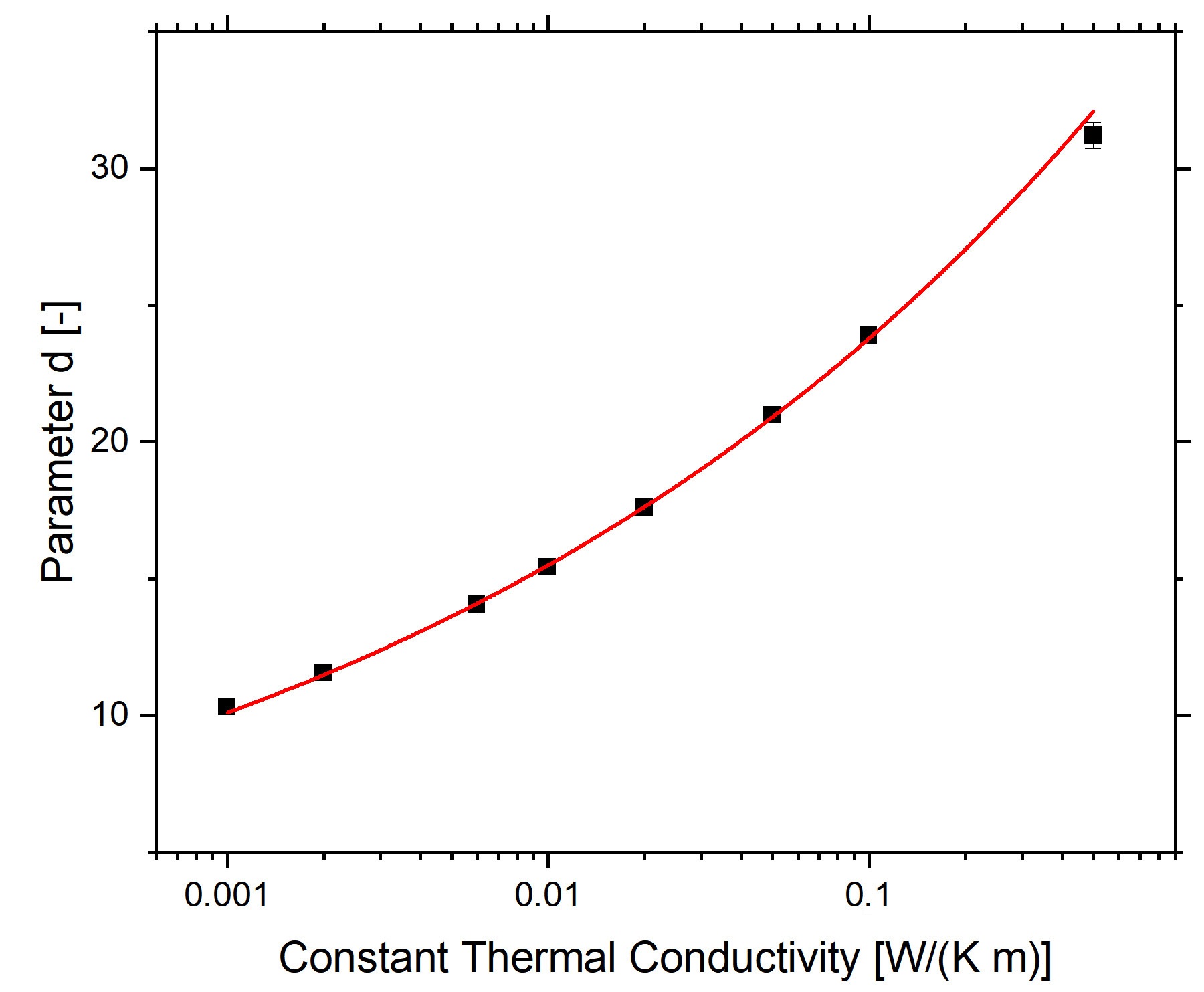}
    \caption{Parameter $d$ (symbols) from Eq. \ref{eq:fit_function_constant} in the micro-porosity case for different constant thermal conductivities $\lambda$. Their relation is approximated by Eq. \ref{eq:parameter_d_const} (solid curve). The goodness of the fit is given by a coefficient of determination of $R_D^2 = 0.999$.}
    \label{fig:Parameter_D_const}
\end{figure}

As can be seen in Figs. \ref{fig:Parameter_C_const} and \ref{fig:Parameter_D_const}, there is also a systematic dependency of the model parameters $c$ and $d$ from Eq. \ref{eq:fit_function_constant} on the thermal conductivity. In order to allow an extrapolation of our model to arbitrary conductivities, we fitted the empirical data with two simple analytic functions of the form
\begin{equation}
    c = 256.18 \, \mathrm{K} \, \left(\frac{\lambda}{\mathrm{1 \, W \, K^{-1}\, m ^{-1}}}\right)^{0.114}
    \label{eq:parameter_c_const}
\end{equation}
and
\begin{equation}
    d = 36.50 \, \mathrm{K} \, \left(\frac{\lambda}{\mathrm{1 \, W \, K^{-1}\, m ^{-1}}}\right)^{0.186}.
    \label{eq:parameter_d_const}
\end{equation}
The solid lines in Figs. \ref{fig:Parameter_C_const} and \ref{fig:Parameter_D_const} show the quality of the two fit functions. 


\newpage
\section{Variation of Rotation Period}
\label{App:rotation_variation}
\begin{figure*}
    \centering
    \includegraphics[width=\textwidth]{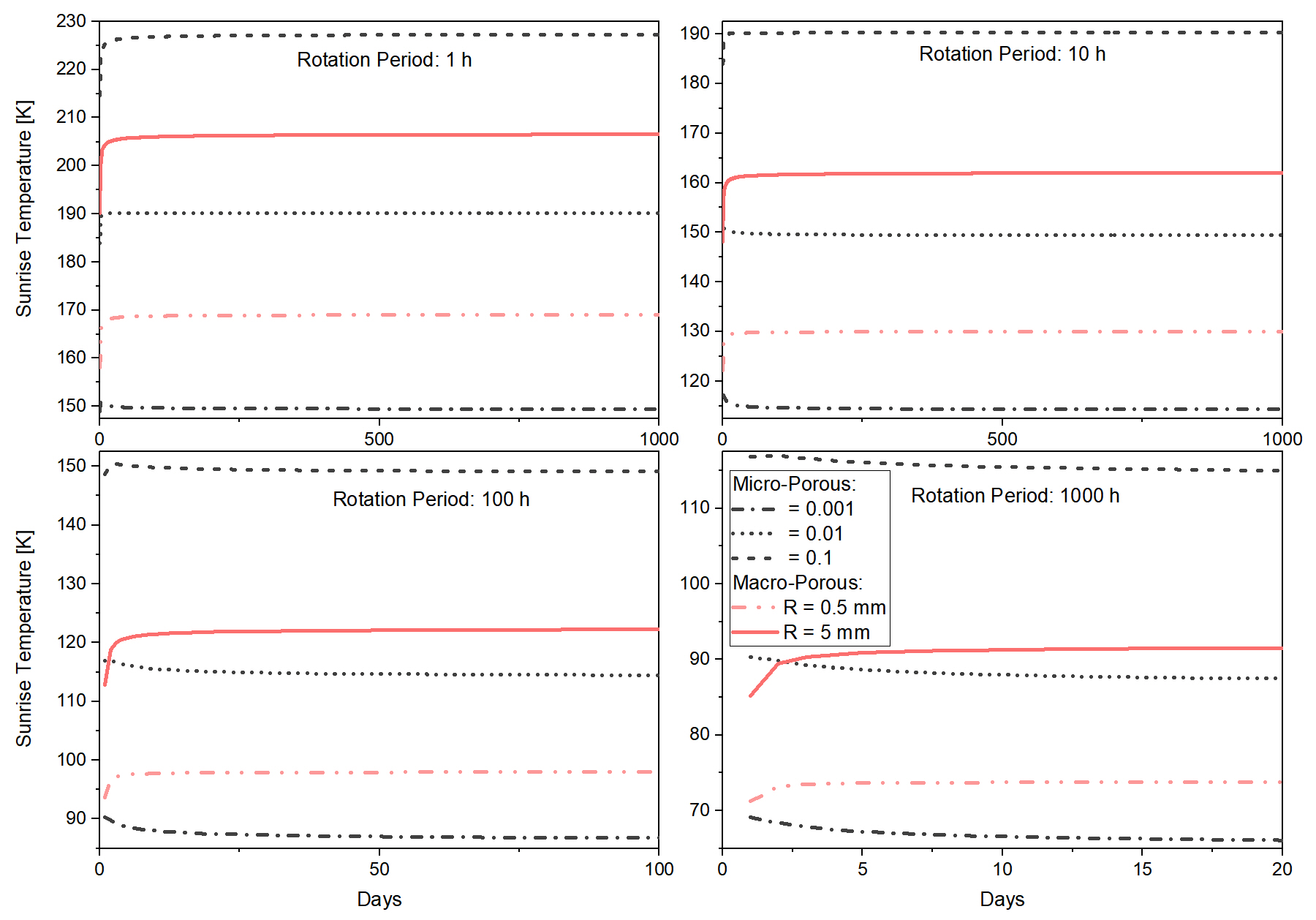}
    \caption{Evolution of surface temperature at sunrise for four different rotation periods as a function of the number of simulation days and for a heliocentric distance of $1.24 \, \mathrm{AU}$. Due to run-time constraints, the cases with spin periods of $100$ hours and $1,000$ hours were run only for $100$ and $20$ days, respectively.}
    \label{fig:rotation_variation_evolution_sunrise_temp}
\end{figure*}

As described in Sec. \ref{sec:results_sub:ideal}, we varied the rotation period to investigate its influence on the sunrise surface temperature. We used rotation periods of $1$, $10$, $100$, and $1,000$ hours, respectively, to cover a wide range of values. Due to run-time constrains, we reduced the simulation duration from $1,000$ comet days to $100$ and $20$ days for the $100$ and $1,000$ hours rotation periods, respectively. To ensure that this reduction has no influence on the comparability, we plotted the evolution of sunrise temperature as a function of the simulation days in Fig. \ref{fig:rotation_variation_evolution_sunrise_temp}. Only during the very first days the values differ from the end of the simulated time. Therefore, comparability is given in all cases. We ran this compatibility check only for the case of $1.24 \, \mathrm{AU}$ heliocentric distance, but we assume that this is also valid for other heliocentric distances.
\par
\begin{figure*}
    \centering
    \begin{minipage}[b]{.47\linewidth}
    \includegraphics[width=\columnwidth]{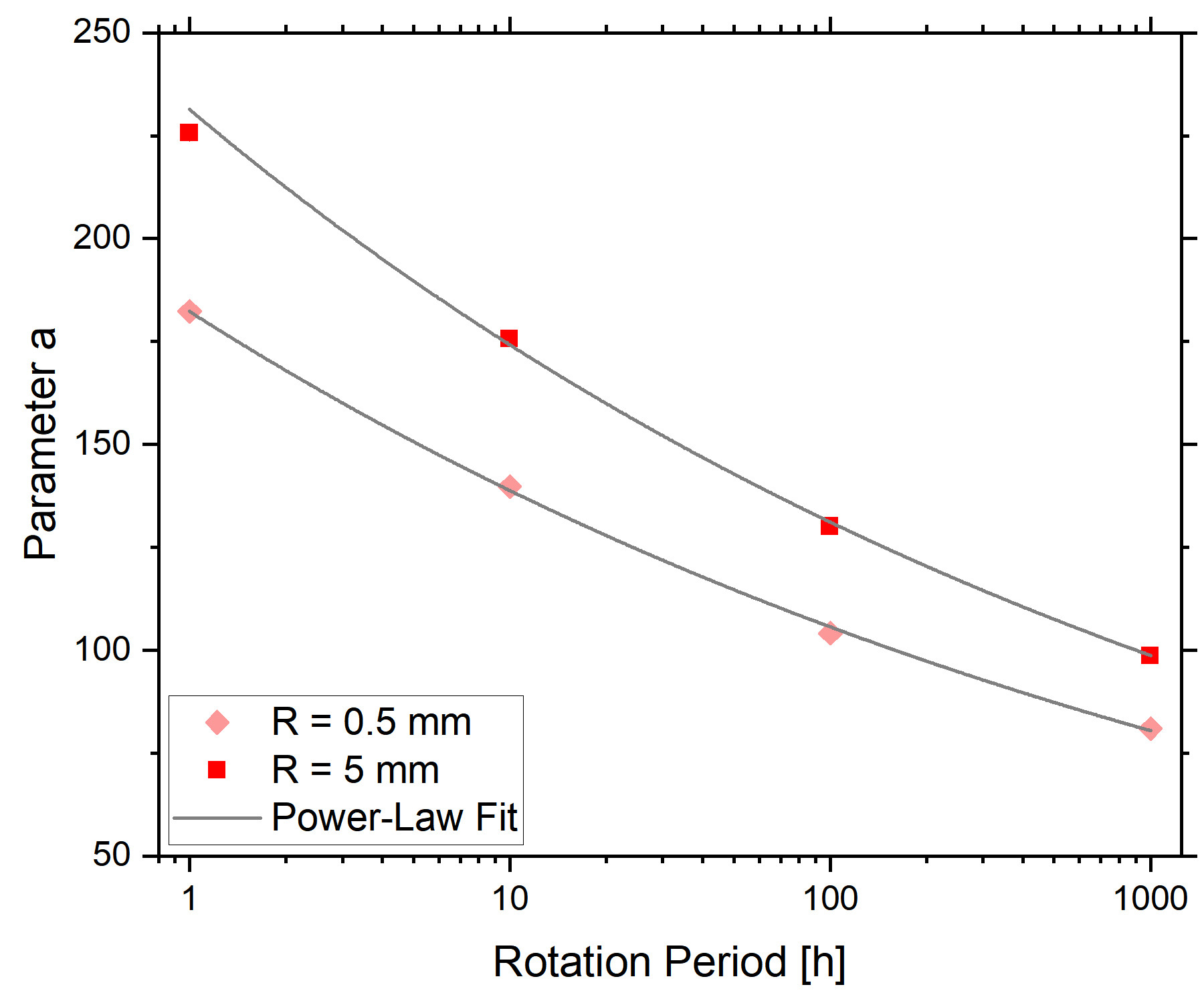}
    \end{minipage}
    \hspace{.05\linewidth}
    \begin{minipage}[b]{.47\linewidth}
    \centering
    \includegraphics[width=\columnwidth]{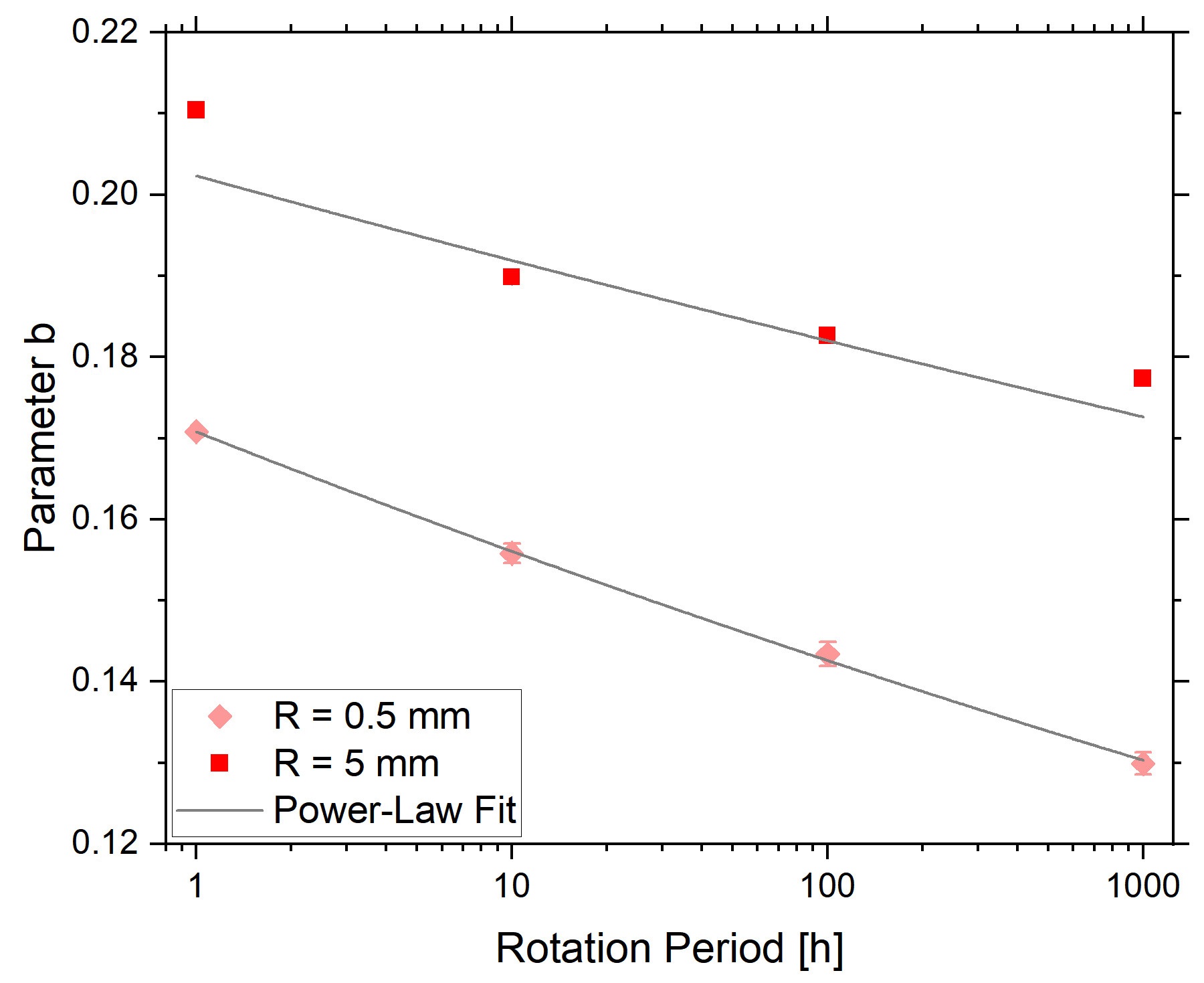}
    \end{minipage}
    \caption{Results for the fit parameters $a$ (left) and $b$ (right) for the macro-porous model with varying rotation periods from Eq. \ref{eq:fit_function_radiative}, see also Table \ref{tab:fit_parameter_radiative_rotation}. A power-law function (Eq. \ref{eq:rotation_variation_parameter_fit}) was fitted to the data. The resulting fit parameters are shown in Table \ref{tab:fit_parameter_rotation_dependency}.}
    \label{fig:Rotation_Pebble_Fitting_Function} 
\end{figure*}
\begin{figure*}
    \centering
    \begin{minipage}[b]{.47\linewidth}
    \includegraphics[width=\columnwidth]{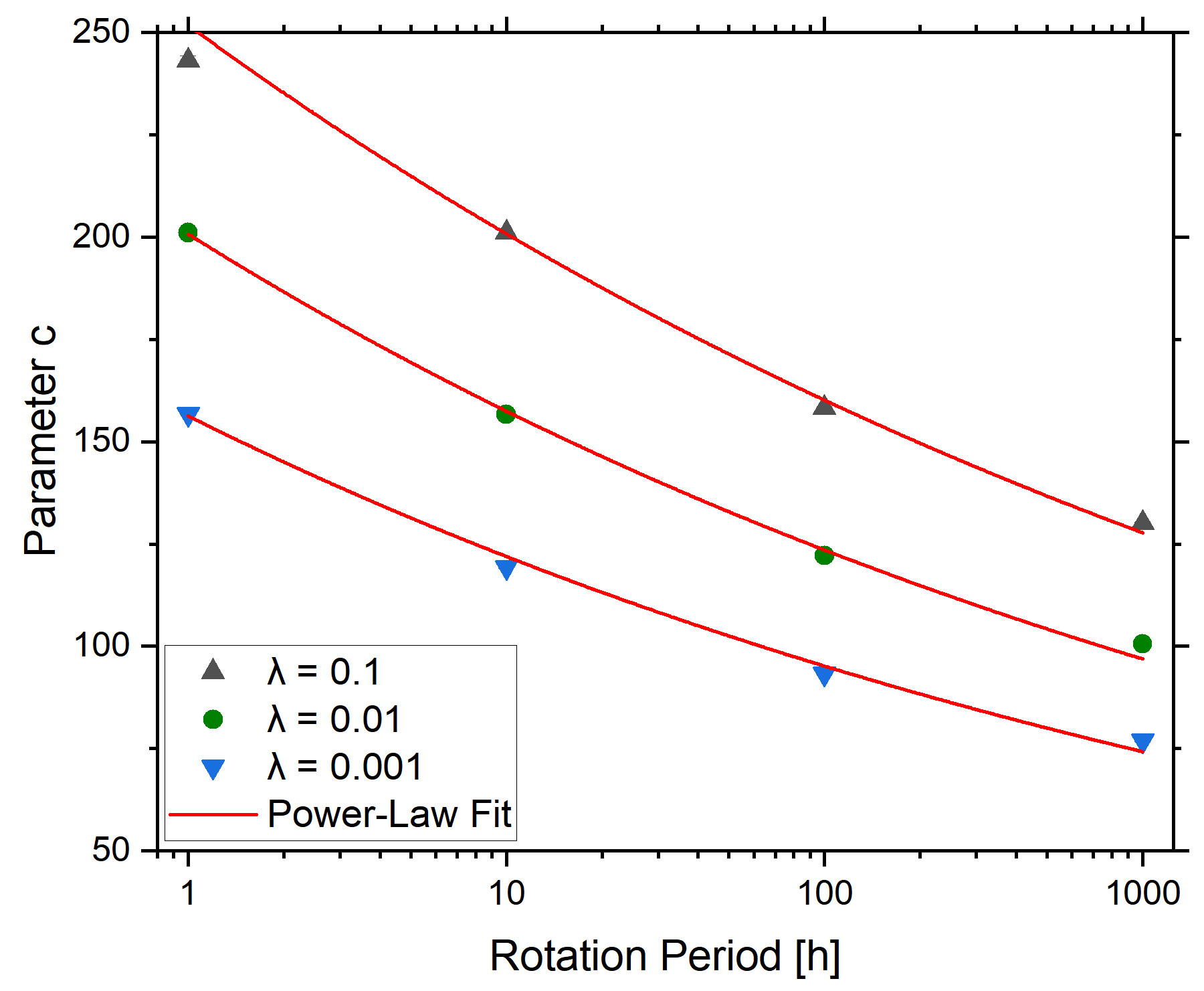}
    \end{minipage}
    \hspace{.05\linewidth}
    \begin{minipage}[b]{.47\linewidth}
    \centering
    \includegraphics[width=\columnwidth]{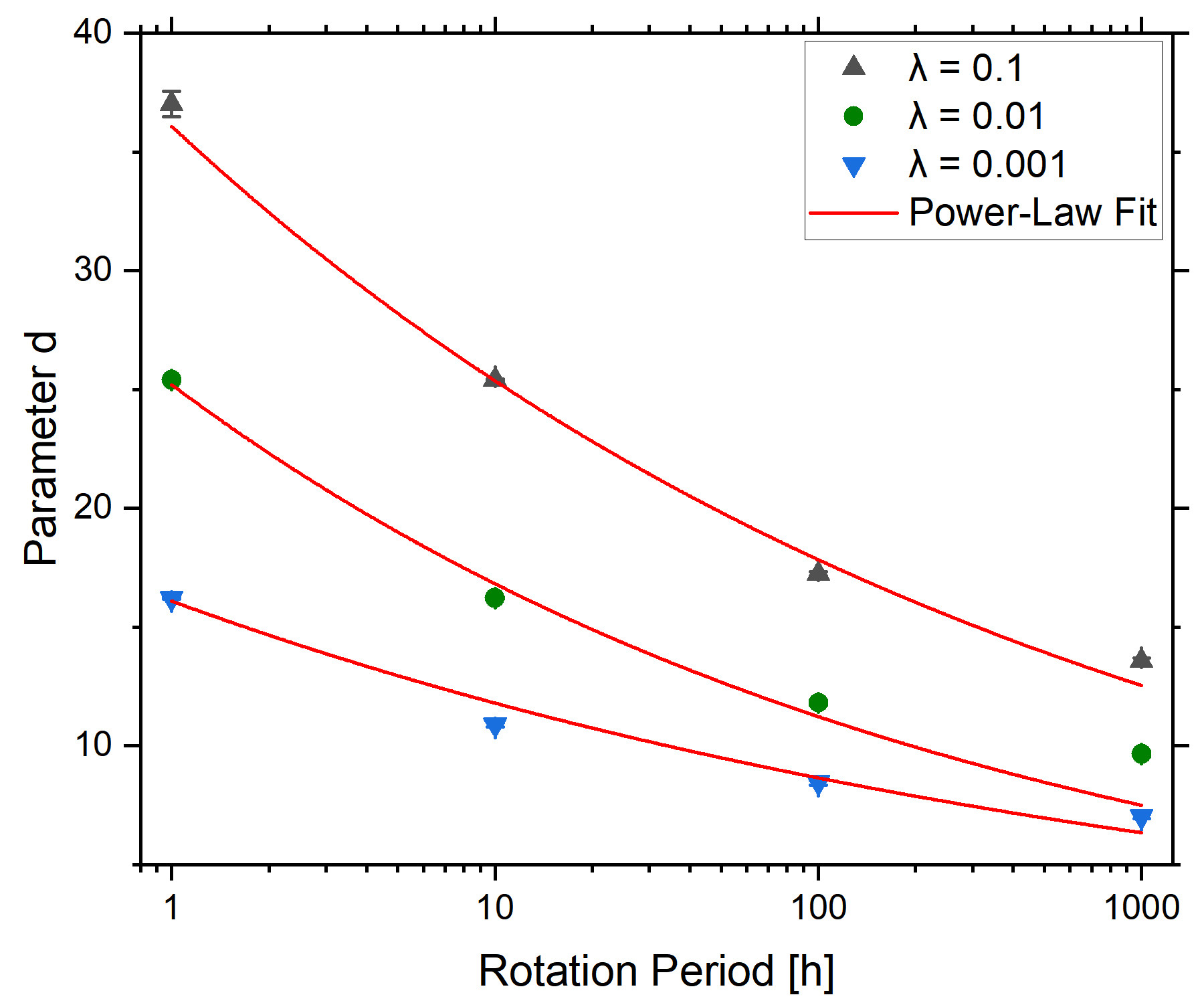}
    \end{minipage}
    \caption{Results for the fit parameters $c$ (left) and $d$ (right) for the micro-porous model with varying rotation periods from Eq. \ref{eq:fit_function_constant}, see also Table \ref{tab:fit_parameter_constant_rotation}. A power-law function (Eq. \ref{eq:rotation_variation_parameter_fit}) was fitted to the data. The resulting fit parameters are shown in Table \ref{tab:fit_parameter_rotation_dependency}.}
    \label{fig:Rotation_Constant_Fitting_Function} 
\end{figure*}

In addition, we fitted the functions Eqs. \ref{eq:fit_function_radiative} and \ref{eq:fit_function_constant} to the sunrise temperatures and retrieved the fit parameters as a function of rotation period. For the macro-porous case, the parameter $a$ and $b$ from Eq. \ref{eq:fit_function_radiative} are given in Table \ref{tab:fit_parameter_radiative_rotation}. The value of the parameters $c$ and $d$ for the micro-porous case from Eq. \ref{eq:fit_function_constant} can be found in Table \ref{tab:fit_parameter_constant_rotation}. Their dependency on rotation period $P$ is plotted in Figs. \ref{fig:Rotation_Pebble_Fitting_Function} and \ref{fig:Rotation_Constant_Fitting_Function}. The data can be fitted by a power-law function
\begin{equation}
    X = \alpha \, P^{\beta} \ ,
    \label{eq:rotation_variation_parameter_fit}
\end{equation}
where $X$ substitutes the parameters $a$, $b$, $c$, or $d$. The resulting fit parameters $\alpha$ and $\beta$ are shown in Table \ref{tab:fit_parameter_rotation_dependency}. In general, the fit quality is high with a coefficient of determination $R_D^2 > 0.98$, with the exception of the $R=5 \, \mathrm{mm}$ pebble case. Here, the fit of parameter $b$ only yields $R_D^2 =0.84$.

\begin{table}
\caption{Fit parameters of the fit function (Eq. \ref{eq:fit_function_radiative}) for the macro-porosity case for varying rotation periods $P$. The quality of the fits are high with coefficients of determination $R_D^2 > 0.998$. }
\centering
\begin{tabular}{cccc}
\hline
Rotation Period & Pebble Radius & Parameter $a$ & Parameter $b$ \\
{[}hours{]} & {[}mm{]}      & {[}K{]}       & {[}-{]}        \\ \hline
1    & 0.5       &  182.3      & 0.171  \\
10   & 0.5       &  139.7      & 0.156  \\
100  & 0.5       &  104.1      & 0.143  \\
1,000 & 0.5       &  80.9       & 0.130  \\
1    & 5         &  225.6      & 0.210  \\
10   & 5         &  175.6      & 0.190  \\
100  & 5         &  130.1      & 0.183  \\
1,000 & 5         &  98.6       & 0.177  \\ \hline     
\end{tabular}%
\label{tab:fit_parameter_radiative_rotation}
\end{table}
\begin{table}
\caption{Fit parameters of the fit function (Eq. \ref{eq:fit_function_constant}) for the micro-porosity case for varying rotation periods $P$. The quality of the fits are high with coefficients of determination $R_D^2 > 0.997$.}
\centering
\begin{tabular}{cccc}
\hline
Rotation Period & Thermal Cond. & Parameter $c$ & Parameter $d$ \\
{[}hours{]} & {[$\mathrm{W/(K \, m)}$]} & {[}K{]}       & {[}K{]}        \\ \hline
1    & 0.1        &  243.2      & 37.0  \\
10   & 0.1        &  201.1      & 25.4  \\
100  & 0.1        &  158.2      & 17.2  \\
1,000 & 0.1        &  130.1      & 13.6  \\
1    & 0.01       &  201.1      & 25.4  \\
10   & 0.01       &  156.7      & 16.2  \\
100  & 0.01       &  122.1      & 11.8  \\
1,000 & 0.01       &  100.5      & 9.7  \\
1    & 0.001      &  156.7      & 16.2  \\
10   & 0.001      &  119.2      & 10.9  \\
100  & 0.001      &  93.2       & 8.4  \\
1,000 & 0.001      &  76.9       & 7.0  \\ \hline    
\end{tabular}%
\label{tab:fit_parameter_constant_rotation}
\end{table}
\begin{table}
\caption{Fit parameters $\alpha$ and $\beta$ from Eq. \ref{eq:rotation_variation_parameter_fit} for the different parameters $a$, $b$, $c$, and $d$ from Eqs. \ref{eq:fit_function_radiative} and \ref{eq:fit_function_constant} as shown in Fig. \ref{fig:Rotation_Pebble_Fitting_Function} and \ref{fig:Rotation_Constant_Fitting_Function} for the simulated macro- and micro-porous cases.}
\centering
\begin{tabular}{ccc}
\hline
Parameter & Parameter $\alpha$ & Parameter $\beta$ \\  \hline
$a$ ($R=0.5\,\mathrm{mm}$)                 &  182.3      & -0.123  \\
$a$ ($R=5\,\mathrm{mm}$)                   &  231.4      & -0.118  \\
$b$ ($R=0.5\,\mathrm{mm}$)                 &  0.171      & -0.039  \\
$b$ ($R=5\,\mathrm{mm}$)                   &  0.202      & -0.023  \\
$c$ ($\lambda = 0.1 \,\mathrm{W/(K \, m)}$)   &  251.8      & -0.098  \\
$c$ ($\lambda = 0.01 \,\mathrm{W/(K \, m)}$)  &  200.7      & -0.105  \\
$c$ ($\lambda = 0.001 \,\mathrm{W/(K \, m)}$) &  156.3      & -0.108  \\
$d$ ($\lambda = 0.1 \,\mathrm{W/(K \, m)}$)   &  36.1      & -0.153  \\
$d$ ($\lambda = 0.01 \,\mathrm{W/(K \, m)}$)  &  25.2      & -0.176  \\
$d$ ($\lambda = 0.001 \,\mathrm{W/(K \, m)}$) &  16.1      & -0.135  \\ \hline    
\end{tabular}%
\label{tab:fit_parameter_rotation_dependency}
\end{table}


\section{Varying The Orbital Scenario: Elliptical Orbit With Obliquity}
\label{App:orbital_scenarios}

\begin{figure*}
    \centering
    \includegraphics[width=\textwidth]{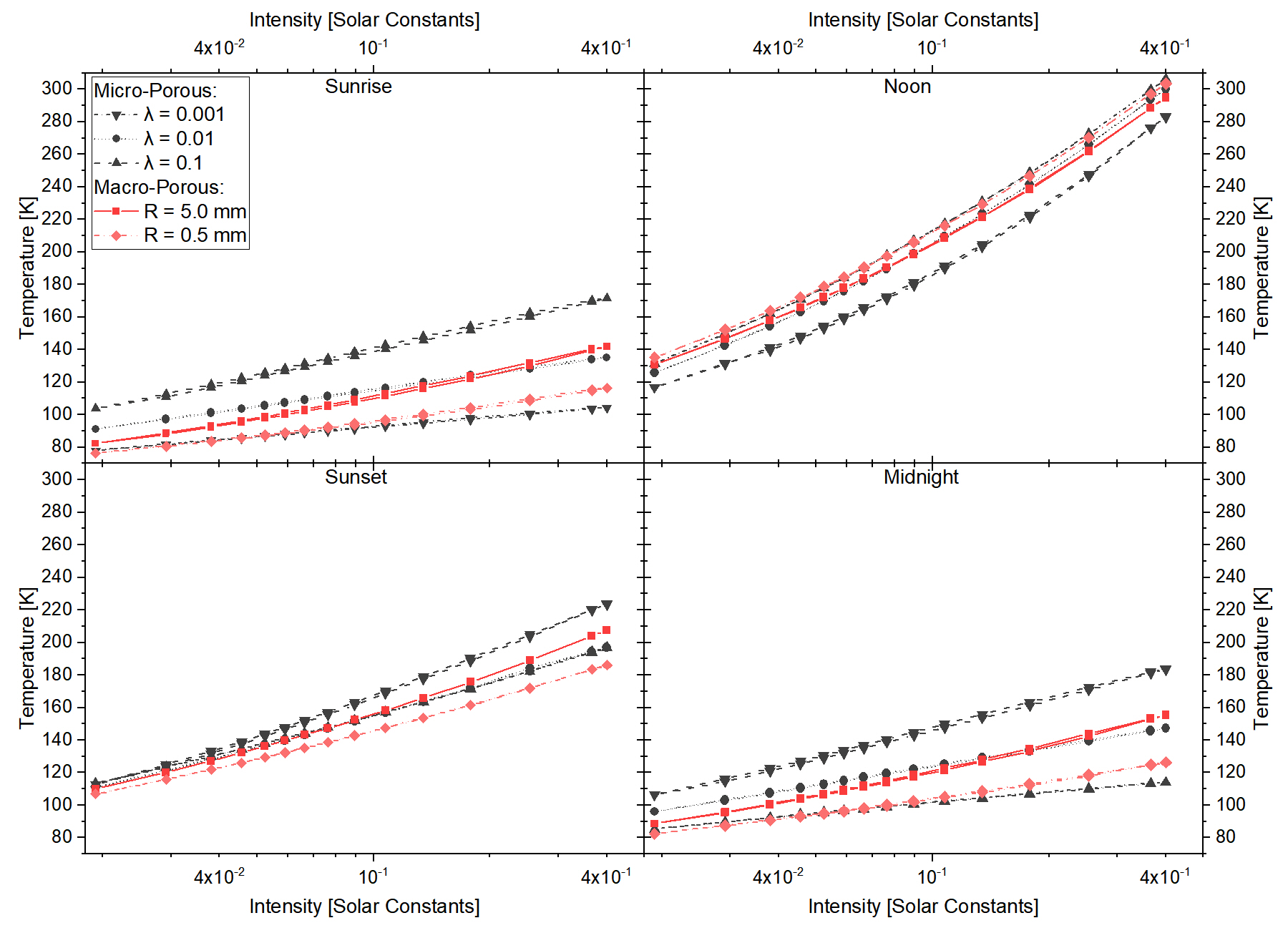}
    \caption{Surface temperature for varying solar intensity for an elliptical orbit with an obliquity of $52^\circ$ at sunrise, noon, sunset and midnight, respectively. The observation point was set on the equator. The macro-porosity cases assume pebble radii of $R = 0.5\, \mathrm{mm}$ (red squares) and $R = 5\, \mathrm{mm}$ (light red diamonds), respectively. The micro-porosity cases (grey) use constant thermal conductivities of $\lambda_2 = 0.001 \, \mathrm{W/(K\, m)}$ (down-pointing triangles), $\lambda_2 = 0.01 \, \mathrm{W/(K\, m)}$ (dots) and $\lambda_2 = 0.1 \, \mathrm{W/(K\, m)}$ (up-pointing triangles), respectively.}
    \label{fig:Axial_Tilt_Temperature_SolarIntensity_Elliptical}
\end{figure*}

In Section \ref{sec:results_sub:realistic}, we introduced a more complex orbital scenario including an elliptical orbit with an axial tilt of $52^\circ$, comparable to the case of comet 67P. Fig. \ref{fig:Axial_Tilt_Temperature_SolarIntensity_Elliptical} shows the dependency of the surface temperature at the equator on the solar intensity for four different times of the day, analogous to Fig. \ref{fig:Circular_Temperature_SolarIntensity} for the circular orbit.


\bsp	
\label{lastpage}
\end{document}